\documentclass[12pt]{iopart}
\usepackage{iopams}
\expandafter\let\csname equation*\endcsname\relax
\expandafter\let\csname endequation*\endcsname\relax
\usepackage{bm,latexsym,amsmath,amssymb,amsfonts,mathrsfs}
\allowdisplaybreaks[1]
\usepackage[dvipdfmx]{hyperref}
\usepackage{url}
\hypersetup{
    colorlinks=true,
    citecolor=cyan,
    linkcolor=magenta,
}
\newcommand*{\D}{{\rm d}}
\newcommand*{\mpl}{M_{\rm Pl}}
\begin{document}

\title[Horndeski theory and beyond]{Horndeski theory and beyond: a review}

\author{Tsutomu~Kobayashi}

\address{Department of Physics, Rikkyo University, Toshima, Tokyo 171-8501, Japan}
\ead{tsutomu@rikkyo.ac.jp}
\vspace{10pt}


\begin{abstract}
This article is intended to review the recent developments
in the Horndeski theory and its generalization,
which provide us with a systematic understanding of
scalar-tensor theories of gravity as well as a powerful tool
to explore astrophysics and cosmology beyond general relativity.
This review covers the generalized Galileons, (the rediscovery of) the
Horndeski theory, cosmological perturbations in the Horndeski theory,
cosmology with a violation of the null energy condition,
degenerate higher-order scalar-tensor theories and their
status after GW170817,
the Vainshtein screening mechanism in the Horndeski theory and beyond,
and hairy black hole solutions.
\end{abstract}

%
%
%
\maketitle
%
%


\section{Introduction}

\subsection{Modified gravity: why?}

General relativity is doubtlessly a very successful theory,
serving as the standard model of gravity.
Nonetheless, modified theories of gravity have been explored
actively for several reasons.

Probably the most major reason in recent years arises from the discovery of
the accelerated expansion of the present universe~\cite{Riess:1998cb,Perlmutter:1998np}.
This may be caused by the (extremely fine-tuned) cosmological constant,
but currently it would be better to have other possibilities at hand
and a long distance modification of general relativity is one of
such possible alternatives. Turning to the accelerated expansion
of the early universe,
it is quite likely that some scalar field called inflaton
provoked inflation~\cite{Starobinsky:1979ty,Guth:1980zm,Sato:1980yn}
and there are a number of models in which the inflaton field is
coupled nonminimally to gravity.
Such inflation models are studied within the context of modified gravity.

In order to test gravity, we need to know predictions of
theories other than general relativity.
This motivation is becoming increasingly important
after the first detection of gravitational waves~\cite{Abbott:2016blz}.
In view of this, modified gravity is worth studying
even if general relativity should turn out to be {\em the}
correct (low-energy effective) description of gravity in the end.

Aside from phenomenology, pursuing consistent modifications of gravity
helps us to learn more deeply about general relativity and gravity.
For example, by trying to develop massive gravity one can
gain a deeper understanding of general relativity
and see how special a massless graviton is. Similarly,
by studying gravity in higher (or lower) dimensions
one can clarify how special gravity in four dimensions is.
This motivation justifies us to study modified gravity
even if we are driven by academic interest.

Finally, one should bear in mind that general relativity is incomplete anyway
as a quantum theory and hence needs to be modified in the UV, though
this subject is beyond the scope of this review.

\subsection{Modified gravity: how?}

Having presented some motivations, let us move to
explain how one can modify general relativity.
According to Lovelock's theorem~\cite{Lovelock:1971yv,Lovelock:1972vz},
the Einstein equations (with a cosmological constant)
are the only possible second-order Euler-Lagrange equations derived
from a Lagrangian scalar density in four dimensions that is constructed
solely from the metric, ${\cal L}={\cal L}[g_{\mu\nu}]$.
To extend Einstein's theory of gravity,
one needs to relax the assumptions of Lovelock's theorem.
The simplest way would be just adding a new degree of freedom
other than the metric, such as a scalar field.
Higher-dimensional gravity may be described by
an effective scalar-tensor theory in four dimensions
via a dimensional reduction.
Incorporating higher derivatives
may lead to a pathological theory (as will be argued shortly)
or something that can be recast in a scalar-tensor theory
(e.g., $R^2$ gravity).
Abandoning diffeomorphism invariance is also equivalent to
introducing new degrees of freedom.
Thus, modifying gravity amounts to changing the degrees of freedom
in any case. In particular,
many different theories of modified gravity can be
described at least effectively
by some additional scalar
degree(s) of freedom on top of the usual two tensor degrees
corresponding to gravitational waves.
We therefore focus on
{\em scalar-tensor theories} in this review.

\subsection{Ostrogradsky instability}

One of the guiding principles we follow
when we
seek for a ``healthy'' extension of general relativity
is to avoid what is called the
Ostrogradsky instability~\cite{Ostrogradsky:1850fid,Woodard:2015zca}.
The theorem states that a system described by
nondegenerate higher-derivative Lagrangian suffers from
ghost-like instabilities.
We will demonstrate this below
by using a simple example in the context of mechanics.

Let us consider the following Lagrangian
involving a second derivative:
\begin{align}
L=\frac{a}{2}\ddot\phi^2-V(\phi),\label{Ost:eq1}
\end{align}
where $a\,(\neq 0)$ is a constant and $V(\phi)$ is an arbitrary potential.
The Euler-Lagrange equation derived from~\eqref{Ost:eq1} is of fourth order:
$a\ddddot\phi-\D V/\D\phi =0$.
To solve this we need four initial conditions, which means that
we have in fact two dynamical degrees of freedom.
According to the Ostrogradsky theorem, one of them must be a ghost.
This can be seen as follows.
By introducing an auxiliary variable,
the Lagrangian~\eqref{Ost:eq1} can be written equivalently as
\begin{align}
L&=a\psi\ddot\phi-\frac{a}{2}\psi^2-V(\phi)
\notag \\
&=-a\dot\psi\dot\phi-\frac{a}{2}\psi^2-V(\phi)+a \frac{\D}{\D t}\left(\psi\dot\phi\right).
\label{Ost:eq2}
\end{align}
It is easy to see that the first line reproduces the original Lagrangian~\eqref{Ost:eq1}
after substituting the Euler-Lagrange equation for $\psi$, namely, $\psi=\ddot\phi$.
The last term in the second line does not contribute to the Euler-Lagrange equation.
In terms of the new variables defined as
$q=(\phi+\psi)/\sqrt{2}$ and $Q=(\phi-\psi)/\sqrt{2}$,
the Lagrangian~\eqref{Ost:eq2} can be rewritten
(up to a total derivative)
in the form
\begin{align}
L=-\frac{a}{2}\dot q^2+\frac{a}{2}\dot Q^2 - U(q,Q).
\end{align}
This Lagrangian clearly shows that the system contains two dynamical degrees of freedom,
one of which has a wrong sign kinetic term,
signaling ghost instabilities.
This is true irrespective of the sign of $a$.

Although we have seen the appearance of the Ostrogradsky instability
in higher-derivative systems only through the above simple example,
this is generically true in higher-derivative field theory.
The theorem can be extended to the systems with
third-order equations of motion~\cite{Motohashi:2014opa}.
In this review, we will therefore consider scalar-tensor modifications of
general relativity that have second-order field equations.
The most general form of the Lagrangian for the
scalar-tensor theory having second-order field equations
has been known as the Horndeski theory~\cite{Horndeski:1974wa},
and it has been widely used in cosmology and astrophysics
beyond general relativity
over recent years.

An important postulate of the Ostrogradsky theorem is
that the Lagrangian is nondegenerate.
If this is not the case, one can reduce a set of
higher-derivative field equations
to a healthy second-order system.
This point will also be discussed
in the context of scalar-tensor theories.

\subsection{Structure of the review}

The outline of this article is as follows.

In the next section,
we review aspects of the Horndeski theory,
the most general scalar-tensor theory with
second-order equations of motion.
It is shown that the original form of the Horndeski action is
indeed equivalent to its modern form frequently used in the literature
(i.e., the generalized Galileons).
A short status report is also given on the attempt to extend
the Horndeski theory to allow for multiple scalar fields.

We then present some applications of the Horndeski theory
to cosmology in Sec.~\ref{sec:cosmoH}.

Scalar-tensor
theories that are more general than Horndeski necessarily have
higher-order equations of motion. Nevertheless, one can circumvent
the Ostrogradsky instability if the system is degenerate, as argued above.
This idea gives rise to new healthy scalar-tensor theories beyond Horndeski.
We review the recent developments in this direction in Sec.~\ref{sec:beyondH}.
The nearly simultaneous detection of gravitational waves GW170817 and the
$\gamma$-ray burst GRB 170817A places a very tight constraint on the speed of
gravitational waves. We mention the status of
the Horndeski theory and its extensions
after this event.

The Vainshtein screening mechanism is essential for
modified gravity evading solar-system constraints.
In Sec.~\ref{sec:Vainshtein_mech}, we describe this mechanism
based on the Horndeski theory and theories beyond Horndeski,
emphasizing in particular
the interesting phenomenology of partial breaking of
Vainshtein screening inside matter
in degenerate higher-order scalar-tensor theories.

Black hole solutions in the Horndeski theory and beyond are
summarized very briefly in Sec.~\ref{sec:BHs}.

Finally, we draw our conclusion in Sec.~\ref{sec:final}.

This review only covers scalar-tensor theories.
For a more comprehensive review
including other types of modified gravity,
see~\cite{Clifton:2011jh,Heisenberg:2018vsk}.
We will not describe much about cosmological tests of gravity,
which are covered by excellent reviews such
as~\cite{Jain:2010ka,Joyce:2014kja,Koyama:2015vza}.

\section{Horndeski theory}

\subsection{From Galileons to Horndeski theory}\label{subsec:GtoH}

To introduce the Horndeski theory in a pedagogical manner,
we start from the Galileon theory
(see also~\cite{Deffayet:2013lga} for a review on the same subject).
The Galileon~\cite{Nicolis:2008in}
is a scalar field with the symmetry under the transformation
$\phi\to \phi + b_\mu x^\mu + c$. This is called,
by an analogy to
a Galilei transformation in classical mechanics,
the Galilean shift symmetry.
In order to avoid ghost instabilities, we demand that $\phi$'s equation of motion
is of second order. The most general Lagrangian
(in four dimensions) having these properties
is given by~\cite{Nicolis:2008in}
\begin{align}
{\cal L}&=c_1\phi +c_2X-c_3X\Box\phi
\notag \\ & \quad
+\frac{c_4}{2}\left\{
X\left[(\Box\phi)^2-\partial_\mu\partial_\nu\phi\partial^\mu\partial^\nu\phi\right]
+\Box\phi\partial^\mu\phi\partial^\nu\phi\partial_{\mu}\partial_{\nu}\phi
-\partial_\mu X\partial^\mu X
\right\}
\notag \\ & \quad
+\frac{c_5}{15}\bigl\{
-2X\left[
(\Box\phi)^3-3\Box\phi\partial_\mu\partial_\nu\phi\partial^\mu\partial^\nu\phi
+2\partial_\mu\partial_\nu\phi\partial^\nu\partial^\lambda\phi
\partial_\lambda\partial^\mu\phi
\right]
\notag \\ & \quad\qquad
+3\partial^\mu\phi\partial_\mu X
\left[(\Box\phi)^2-\partial_\mu\partial_\nu\phi\partial^\mu\partial^\nu\phi\right]
+6\Box\phi \partial_\mu X\partial^\mu X-6 \partial^\mu\partial^\nu\phi
\partial_\mu X\partial_\nu X
\bigr\}
,\label{eq:Lag_original_Galileon1}
\end{align}
where $X:=-\eta^{\mu\nu}\partial_\mu\phi\partial_\nu\phi/2$
and $c_1,\cdots,c_5$ are constants.
This can be written in a more compact form by
making use of integration by parts as
\begin{align}
{\cal L}&=c_1\phi +c_2X-c_3X\Box\phi+c_4 X
\left[(\Box\phi)^2-\partial_\mu\partial_\nu\phi\partial^\mu\partial^\nu\phi\right]
\notag \\ & \quad
-\frac{c_5}{3}X\left[
(\Box\phi)^3-3\Box\phi\partial_\mu\partial_\nu\phi\partial^\mu\partial^\nu\phi
+2\partial_\mu\partial_\nu\phi\partial^\nu\partial^\lambda\phi
\partial_\lambda\partial^\mu\phi
\right].\label{eq:Lag_original_Galileon2}
\end{align}
Note that the field equation is of
second order even though the Lagrangian depends on the
second derivatives of the field.

The Lagrangian~\eqref{eq:Lag_original_Galileon2}
describes a scalar-field theory on a fixed Minkowski background.
One can introduce gravity and consider a covariant version
of~\eqref{eq:Lag_original_Galileon2}
by promoting $\eta_{\mu\nu}$ to $g_{\mu\nu}$ and $\partial_\mu$ to $\nabla_\mu$.
However, since covariant derivatives do not commute,
the naive covariantization leads to higher derivatives
in the field equations, which would be dangerous.
For example,
one would have derivatives of the Ricci tensor $R_{\mu\nu}$
from the term having the coefficient $c_4$,
\begin{align}
c_4 X\nabla^\mu \left[
\nabla_\mu \nabla_\nu\nabla^\nu\phi -
\nabla_\nu \nabla_\mu \nabla^\nu \phi
\right]
=- c_4 X\nabla^\mu\left(R_{\mu\nu}\nabla^\nu\phi\right),
\end{align}
in the scalar-field equation of motion.
Such higher derivative terms can be canceled
by adding curvature-dependent terms
appropriately to Eq.~\eqref{eq:Lag_original_Galileon2}.
The covariant version of~\eqref{eq:Lag_original_Galileon2}
that leads to second-order field equations
both for the scalar field and the metric is given by~\cite{Deffayet:2009wt}
\begin{align}
  {\cal L}&=c_1\phi +c_2X-c_3X\Box\phi
+\frac{c_4}{2}X^2R
  +c_4 X
  \left[(\Box\phi)^2-\phi^{\mu\nu}\phi_{\mu\nu}\right]
  \notag \\ & \quad
  +c_5X^2 G^{\mu\nu}\phi_{\mu\nu}
  -\frac{c_5}{3}X\left[
  (\Box\phi)^3-3\Box\phi\phi^{\mu\nu}\phi_{\mu\nu}
  +2\phi_{\mu\nu}\phi^{\nu\lambda}\phi_\lambda^\mu
  \right],\label{eq:Lag_cov_Galileon}
\end{align}
where $R$ is the Ricci tensor, $G_{\mu\nu}$ is the Einstein tensor,
$\phi_\mu:=\nabla_\mu\phi$,
$\phi_{\mu\nu}:=\nabla_\mu\nabla_\nu\phi$
and now $X:=-g^{\mu\nu}\phi_\mu\phi_\nu/2$.
Here, the fourth term in the first line and the first term in the second line
are the ``counter terms''
introduced to remove higher derivatives in the field equations.
The counter terms are unique.
This theory is called the covariant Galileon.
Since the field equations derived from the Lagrangian~\eqref{eq:Lag_cov_Galileon}
involve first derivatives of $\phi$,
the Galilean shift symmetry is broken in the covariant Galileon theory.
Only the second property of the Galileon, i.e.,
the second-order nature of the field equations,
is maintained in the course of covariantization.
The covariant Galileon theory~\eqref{eq:Lag_cov_Galileon}
is formulated in four spacetime dimensions, but it
can be extended to arbitrary dimensions~\cite{Deffayet:2009mn}.

The generalized Galileon~\cite{Deffayet:2011gz}
is a further generalization of the covariant Galileon~\cite{Deffayet:2009wt,Deffayet:2009mn}
retaining second-order field equations.
More precisely,
first one determines the most general scalar-field theory
on a fixed Minkowski background which yields a second-order field equation,
assuming that
the Lagrangian contains at most second derivatives of $\phi$
and is polynomial in $\partial_\mu\partial_\nu\phi$.
One then promotes the theory to
a covariant one in the same way as above by adding
appropriate (unique) counter terms so that the
field equations are of second order both for $\phi$ and the metric.
The generalized Galileon can thus be obtained.
It should be noted that this procedure can be done in
arbitrary spacetime dimensions.
In four dimensions, the Lagrangian
for the generalized Galileon
is given by~\cite{Deffayet:2011gz}
\begin{align}
{\cal L}&=G_2(\phi,X)-G_3(\phi,X)\Box\phi + G_4(\phi,X)R
+G_{4X}  \left[(\Box\phi)^2-\phi^{\mu\nu}\phi_{\mu\nu}\right]
  \notag \\ & \quad
  +G_5(\phi,X) G^{\mu\nu}\phi_{\mu\nu}
  -\frac{G_{5X}}{6}\left[
  (\Box\phi)^3-3\Box\phi\phi^{\mu\nu}\phi_{\mu\nu}
  +2\phi_{\mu\nu}\phi^{\nu\lambda}\phi_\lambda^\mu
  \right],\label{eq:Lag_gen_Galileon}
\end{align}
where $G_2$, $G_3$, $G_4$, and $G_5$ are arbitrary functions of $\phi$ and $X$.
Here and hereafter we use the notation $f_X:=\partial f/\partial X$
and $f_\phi:=\partial f/\partial \phi$ for a function $f$ of $\phi$ and $X$.

The generalized Galileon~\eqref{eq:Lag_gen_Galileon} is now known as
the Horndeski theory~\cite{Horndeski:1974wa}, i.e.,
{\em the most general scalar-tensor theory having second-order
field equations in four dimensions}.
However, Horndeski determined the theory
starting from the different assumptions than those made for deriving
the generalized Galileon, and the original form of the Lagrangian~\cite{Horndeski:1974wa}
looks very different from~\eqref{eq:Lag_gen_Galileon}:
\begin{align}
{\cal L}&=\delta_{\mu\nu\sigma}^{\alpha\beta\gamma}
\left[
\kappa_1  \phi^\mu_\alpha R_{\beta \gamma}^{~~\;\nu\sigma}+\frac{2}{3}\kappa_{1X}
\phi^\mu_\alpha\phi^\nu_\beta\phi^\sigma_\gamma
+\kappa_3\phi_\alpha\phi^\mu R_{\beta\gamma}^{\;~~\nu\sigma}
+2\kappa_{3 X}\phi_\alpha \phi^\mu \phi^\nu_\beta \phi^\sigma_\gamma
\right]
\notag \\ & \quad
+\delta^{\alpha\beta}_{\mu\nu}\left[
(F+2W)R_{\alpha\beta}^{\;~~ \mu\nu}+2F_X\phi^\mu_\alpha \phi^\nu_\beta
+2\kappa_8\phi_\alpha\phi^\mu\phi^\nu_\beta\right]
\notag \\ &\quad
-6\left(F_\phi+2W_\phi-X\kappa_8\right)\Box\phi+\kappa_9.
\label{eq:Lag_original_Horndeski}
\end{align}
Here,
$\delta^{\alpha_1\alpha_2...\alpha_n}_{\mu_1\mu_2...\mu_n}:=n!\delta_{\mu_1}^{[\alpha_1}
\delta_{\mu_2}^{\alpha_2}...\delta_{\mu_n}^{\alpha_n]}$
is the generalized Kronecker delta, and
$\kappa_1$, $\kappa_3$, $\kappa_8$, and $\kappa_9$ are
arbitrary functions of $\phi$ and $X$.
We have another function $F=F(\phi,X)$,
but this must satisfy $F_X=2\left(\kappa_3+2X\kappa_{3X}-\kappa_{1\phi}\right)$
and hence is not independent.
We also have a function of $\phi$, $W=W(\phi)$, which can be absorbed
into the redefinition of $F$: $F_{\rm old}+2W\to F_{\rm new}$.
Thus, we have the same number of free functions of $\phi$ and $X$
as in the generalized Galileon theory.
Nevertheless, the equivalence between the two theories is apparently
far from trivial.

In~\cite{Kobayashi:2011nu} it was shown that
the generalized Galileon can be mapped to the Horndeski theory
by identifying $G_i(\phi,X)$ as
\begin{align}
  G_2&=\kappa_9+4X\int^X\D X'\left(\kappa_{8\phi}-2\kappa_{3\phi\phi}\right),
  \\
  G_3&=
  6F_\phi-2X\kappa_8-8X\kappa_{3\phi}+2\int^X\D X'(\kappa_8-2\kappa_{3\phi}),
  \\
  G_4&=2F-4X\kappa_3,
  \\
  G_5&=-4\kappa_1,
\end{align}
and performing integration by parts.
Having thus proven that the generalized Galileon is indeed equivalent
to the Horndeski theory, we can now use~\eqref{eq:Lag_gen_Galileon}
as the Lagrangian for the most general scalar-tensor theory
with second-order field equations.
However,
while the generalized Galileon is formulated
in arbitrary dimensions,
the higher-dimensional extension of the Horndeski theory
has not been known so far and
it is unclear whether or not the generalized Galileon
gives the most general second-order scalar-tensor
theory in higher dimensions.
Note in passing that the lower-dimensional version
of the Horndeski theory
can be obtained straightforwardly~\cite{Horndeski:1974wa}.

The Horndeski theory was obtained already in 1974, but
the paper~\cite{Horndeski:1974wa} had long been forgotten until 2011 when
it was rediscovered by~\cite{Charmousis:2011bf}.
Let us sketch (very briefly) the original
derivation of the Horndeski theory.
The starting point is a generic action of the form
\begin{align}
S=\int \D^4x\sqrt{-g}{\cal L}(g_{\mu\nu}, g_{\mu\nu,\lambda_1},\cdots,
g_{\mu\nu,\lambda_1,\cdots,\lambda_p}, \phi, \phi_{,\lambda_1},
\cdots, \phi_{,\lambda_1,\cdots,\lambda_{q}} ),
\end{align}
with $p,q\ge 2$ {\em in four dimensions}.
The assumptions here should be contrasted with those in the generalized Galileon:
Horndeski's derivation starts from the more general Lagrangian,
but it is restricted to four dimensions.
Varying the action with respect to the metric and the scalar field,
we obtain the field equations:
${\cal E}_{\mu\nu}\left(:=2(\sqrt{-g})^{-1}\delta S/\delta g^{\mu\nu}\right)=0$
and ${\cal E}_\phi\left(:=(\sqrt{-g})^{-1}\delta S/\delta \phi\right)=0$,
where ${\cal E}_{\mu\nu}$ and ${\cal E}_\phi$ are assumed to involve
at most second derivatives of $g_{\mu\nu}$ and $\phi$.
As a consequence of the diffeomorphism invariance of the action,
the following ``Bianchi identity'' holds:
\begin{align}
\nabla^\nu {\cal E}_{\mu\nu}=-\nabla_\mu\phi\, {\cal E}_\phi.
\label{eq:generalized_Bianchi_id}
\end{align}
In general, $\nabla^\nu {\cal E}_{\mu\nu}$ would be of third order
in derivatives of $g_{\mu\nu}$ and $\phi$. However,
since the right-hand side contains at most second derivatives,
$\nabla^\nu {\cal E}_{\mu\nu}$ must be
of second order even though ${\cal E}_{\mu\nu}$ itself is of second order.
This puts a tight restriction on the structure of ${\cal E}_{\mu\nu}$.
Our next step is to construct the tensor ${\cal A}_{\mu\nu}$
satisfying this property. After a lengthy procedure one can
determine the general form of ${\cal A}_{\mu\nu}$ in the end.
(At this step the assumption on the number of spacetime dimensions is used.)
Then, one further restricts the form of ${\cal A}_{\mu\nu}$
by requiring that $\nabla^\nu {\cal A}_{\mu\nu}$ is proportional
to $\nabla_\mu\phi$ as implied by Eq.~\eqref{eq:generalized_Bianchi_id}.
The tensor ${\cal A}_{\mu\nu}$ thus obtained will be ${\cal E}_{\mu\nu}$.
The final step is to seek for the Lagrangian ${\cal L}$ that yields
the Euler-Lagrange equations ${\cal E}_{\mu\nu}=0$ and ${\cal E}_\phi=0$.
Fortunately enough, it turns out that the Euler-Lagrange equations
derived from ${\cal L}=g^{\mu\nu}{\cal E}_{\mu\nu}$
reproduces the structure of ${\cal E}_{\mu\nu}$ and ${\cal E}_\phi$.
This is how we can arrive at the Lagrangian~\eqref{eq:Lag_original_Horndeski}.

By taking the functions in Eq.~\eqref{eq:Lag_gen_Galileon}
appropriately, one can reproduce any second-order scalar-tensor theory
as a specific case.
For example, nonminimal coupling of the form $f(\phi)R$
can be obtained by taking $G_4=f(\phi)$, and
its limiting case $G_4={\rm const}=\mpl^2/2$
gives the Einstein-Hilbert term.
Clearly, $G_2$ is the familiar term used in
k-inflation~\cite{ArmendarizPicon:1999rj}/k-essence~\cite{Chiba:1999ka,ArmendarizPicon:2000dh},
and the $G_3$ term was investigated more recently in the context of
kinetic gravity braiding~\cite{Deffayet:2010qz}/G-inflation~\cite{Kobayashi:2010cm}.
It is well known that $f(R)$ gravity (a theory whose Lagrangian is
given by some function of the Ricci scalar) can be expressed equivalently
as a second-order scalar-tensor theory and hence is a subclass of the
Horndeski theory (see, e.g.,~\cite{Sotiriou:2008rp,DeFelice:2010aj}).
Nonminimal coupling of the form $G^{\mu\nu}\phi_\mu\phi_\nu$
has been studied often in the literature (see, e.g.,~\cite{Germani:2010gm}),
and this term can be expressed in two ways,
$G_4=X$ or $G_5=-\phi$,
with integration by parts.

A nontrivial and interesting example is nonminimal coupling
to the Gauss-Bonnet term,
\begin{align}
  \xi(\phi)\left(R^2-4R_{\mu\nu}R^{\mu\nu}
  +R_{\mu\nu\rho\sigma}R^{\mu\nu\rho\sigma}\right).\label{eq:nonminimal_GB}
\end{align}
No similar terms can be found in~\eqref{eq:Lag_gen_Galileon}
or~\eqref{eq:Lag_original_Horndeski},
but since the Horndeski theory is the most general scalar-tensor
theory with second-order field equations and
it is known that the term~\eqref{eq:nonminimal_GB}
yields the second-order field equations, this {\em must} be obtained
somehow as a specific case of the Horndeski theory.
In fact, one can reproduce~\eqref{eq:nonminimal_GB} by taking~\cite{Kobayashi:2011nu}
\begin{align}
  &G_2=8\xi^{(4)}X^2\left(3-\ln X\right),
  \quad  G_3=4\xi^{(3)}X\left(7-3\ln X\right),
  \notag \\
  &G_4=4\xi^{(2)}X\left(2-\ln X\right),
  \quad\;\; G_5=-4\xi^{(1)} \ln X,\label{eq:GB_Galileon_expression}
\end{align}
where $\xi^{(n)}:=\partial^n\xi/\partial\phi^n$.
To confirm that~\eqref{eq:GB_Galileon_expression} is
indeed equivalent to~\eqref{eq:nonminimal_GB}
at the level of the action is probably extremely difficult,
but it is straightforward to see the equivalence if one works at the level of
the equations of motion.
Note in passing that a function of
the Gauss-Bonnet term in a Lagrangian,
$f(R^2-4R_{\mu\nu}R^{\mu\nu}+R_{\mu\nu\rho\sigma}R^{\mu\nu\rho\sigma})$,
can be recast into the form of~\eqref{eq:nonminimal_GB} by introducing
an auxiliary field, and hence it is also included in the Horndeski theory.

Another nontrivial example is the derivative coupling to the
double dual Riemann tensor,
\begin{align}
\phi_\mu\phi_\nu\phi_{\alpha\beta}L^{\mu\alpha\nu\beta},
\end{align}
where
\begin{align}
L^{\mu\alpha\nu\beta}:=R^{\mu\alpha\nu\beta}+
\left(R^{\mu\beta}g^{\nu\alpha}+R^{\nu\alpha}g^{\mu\beta}
-R^{\mu\nu}g^{\alpha\beta}-R^{\alpha\beta}g^{\mu\nu}\right)
+\frac{1}{2}R\left(g^{\mu\nu}g^{\alpha\beta}-g^{\mu\beta}g^{\nu\alpha}\right).
\end{align}
This can be reproduced simply by taking $G_5=X$~\cite{Narikawa:2013pjr}.

There are other well-motivated models or scenarios
which have some links to the Galileon/Horndeski theory.
The Dvali-Gabadadze-Porrati (DGP) model~\cite{Dvali:2000hr} based on
a five-dimensional braneworld scenario gives rise to
the cubic Galileon interaction $\sim (\partial\phi)^2\Box\phi$
in its four-dimensional effective theory~\cite{Luty:2003vm}.
Actually, the Galileon was originally proposed as a
generalization of the DGP effective theory.
The Dirac-Born-Infeld (DBI) action,
which is described by a particular form of $G_2(\phi,X)$ and
often studied in the context of inflation~\cite{Silverstein:2003hf,Alishahiha:2004eh},
can be obtained from a probe brane moving in a five-dimensional bulk spacetime.
By extending the probe brane action,
one can similarly derive the generalization of the Galileon whose action
is of the particular Horndeski form involving
$G_3$, $G_4$, and $G_5$~\cite{deRham:2010eu,Goon:2011qf,Goon:2011uw,Trodden:2011xh}.
It is shown in~\cite{VanAcoleyen:2011mj}
and revisited in~\cite{vandeBruck:2018jlz} that
some particular cases of the Horndeski action can be obtained
through a Kaluza-Klein compactification
of higher-dimensional Lovelock gravity.
The nonminimal couplings in the Horndeski theory
capture the essential structure of (the decoupling limit of)
massive gravity~\cite{deRham:2011by,Heisenberg:2014kea}.

\subsection{ADM decomposition}

For the later purpose, it is convenient to perform a $3+1$
Arnowitt-Deser-Misner (ADM) decomposition~\cite{Arnowitt:1962hi}
in the Horndeski theory.
We take the unitary gauge
in which $\phi$ is homogeneous on constant-time hypersurfaces,
so that $\phi=\phi(t)$. (We assume that it is possible to take such
a coordinate system.)
The metric can be expressed as
\begin{align}
\D s^2=-N^2\D t^2+\gamma_{ij}\left(\D x^i+N^i\D t\right)\left(\D x^j+N^j\D t\right),
\end{align}
where $N$ is the lapse function, $N_i\;\left(=\gamma_{ij}N^j\right)$
is the shift vector, and $\gamma_{ij}$ is the three-dimensional spatial metric.
Then, since $X=\dot\phi^2(t)/(2N^2)$,
the functions of $\phi$ and $X$ can be regarded as those of $t$ and $N$.
We will need the extrinsic curvature of the spatial hypersurfaces,
\begin{align}
K_{ij}:=\frac{1}{2N}\left(\dot\gamma_{ij}-D_iN_j-D_jN_i\right),
\end{align}
where a dot denotes differentiation with respect to $t$ and
$D_i$ is the covariant derivative associated with $\gamma_{ij}$.
The second derivatives of $\phi$ can be expressed using $K_{ij}$.
For example, we have $\phi_{ij}=-(\dot\phi/N)K_{ij}$.
The four-dimensional Ricci tensor can also be expressed
using the extrinsic curvature and the three-dimensional
Ricci tensor, $R_{ij}^{(3)}$. With some manipulation,
we find that the Horndeski action in the ADM form is
given by~\cite{Gleyzes:2014dya}
\begin{align}
S=\int \D t\D^3x \sqrt{\gamma} N& \biggl[
A_2(t,N)+A_3(t,N)K
+B_4(t,N)R^{(3)}
\notag \\ &
-(B_4+NB_{4N})\left(K^2-K_{ij}K^{ij}\right)
+B_5(t,N)G_{ij}^{(3)}K^{ij}
\notag \\ &
+\frac{NB_{5N}}{6}\left(
K^3-3KK_{ij}K^{ij}+2K_{ij}K^{jk}K_k^i
\right)
\biggr].\label{action:ADM_form}
\end{align}
Now we have four free functions of $t$ and $N$,
which are related to $G_2$, $G_3$, $G_4$, and $G_5$ as
\begin{align}
A_2&=G_2+\sqrt{X}\int^X\frac{G_{3\phi}}{\sqrt{X'}}\D X',
\\
A_3&=\int^X G_{3X'}\sqrt{2X'}\D X'-2\sqrt{2X}G_{4\phi},
\\
B_4&=G_4-\frac{\sqrt{X}}{2}\int^X\frac{G_{5\phi}}{\sqrt{X'}}\D X',
\\
B_5&=-\int^X G_{5X'}\sqrt{2X'}\D X'.
\end{align}
The above ADM form of the action is particularly useful for
studying cosmology in the Horndeski theory.

Since the scalar field is apparently gone in the ADM description,
one might wonder how one can understand from the action~\eqref{action:ADM_form}
that the theory has $(2+1)$ dynamical degrees of freedom.
The point is that $\delta S/\delta N=0$ gives the equation
that determines $N$ in terms of $\gamma_{ij}$ and $\dot\gamma_{ij}$
rather than a constraint among $\gamma_{ij}$ and $\dot\gamma_{ij}$,
which signals an extra degree of freedom.
Note that this remains true even if one generalizes the ADM action to
\begin{align}
S=\int\D t\D^3x \sqrt{\gamma}N&\bigl[\cdots +
B_4(t,N)R^{(3)}+C_4(t,N)\left(K^2-K_{ij}K^{ij}\right)
\notag \\  &
+B_5(t,N)G_{ij}^{(3)}K^{ij}+C_5(t,N)\left(K^3+\cdots\right)
\bigr],
\end{align}
where $B_4$, $B_5$, $C_4$, and $C_5$ are {\em independent} functions.
This idea hints at the possibility of generalizing the Horndeski theory
while retaining the number of dynamical degrees of freedom.
Indeed, the Gleyzes-Langlois-Piazza-Vernizzi (GLPV)
generalization of the Horndeski theory was
noticed in this way~\cite{Gleyzes:2014dya,Gleyzes:2014qga}.
We will come back to the GLPV theory in Sec.~\ref{sec:dhost}.
See also Refs.~\cite{Gao:2014soa,Gao:2014fra,Saitou:2016lvb,Lin:2017utd,Gao:2018znj}
for a further generalization
of the ADM description of scalar-tensor theories.

\subsection{Multi-scalar generalization}

Having determined the most general {\em single}-scalar-tensor theory
with second-order field equations,
it is natural to explore its multi-scalar generalization.
However, so far no complete multi-scalar version of the Horndeski theory
has been obtained. Let us summarize the current status of
attempts to generalize the Galileon/Horndeski theory to
multiple scalar fields.

The Galileon is generalized to mixed combinations of $p$-form fields
in~\cite{Deffayet:2010zh}, a special case of which is the
bi- and multi- Galileon theory~\cite{Padilla:2010de,Padilla:2010tj}.
The multi-Galileon theory can also be derived from
a probe brane embedded in a higher-dimensional bulk
by extending the method of~\cite{deRham:2010eu}
to the case with higher co-dimensions~\cite{Trodden:2011xh,Hinterbichler:2010xn}.
The multi-Galileon theory can be promoted to involve
arbitrary functions of the $N$ scalar fields $\phi^I$ ($I=1,\cdots, N$)
and their kinetic terms
$-\partial_\mu\phi^I\partial^\mu\phi^J/2$~\cite{Padilla:2012dx,Sivanesan:2013tba}.
Similarly to the single-field Galileon,
the multi-Galileon can be covariantized,
while maintaining the second-order nature, to give~\cite{Padilla:2012dx}
\begin{align}
{\cal L}&=G_2-G_{3I}\Box\phi^I +G_4 R+G_{4,\langle IJ\rangle}
\left(\Box\phi^I\Box\phi^J-\nabla_\mu\nabla_\nu\phi^I\nabla^\mu\nabla^\nu\phi^J\right)
\notag \\ &\quad
+G_{5I}G^{\mu\nu}\nabla_\mu\nabla_\nu\phi^I
-\frac{1}{6}G_{I,\langle JK\rangle }\bigl[
\Box\phi^I\Box\phi^J\Box\phi^K-
3\Box\phi^{(I}\nabla_\mu\nabla_\nu\phi^J\nabla^\mu\nabla^\nu\phi^{K)}
\notag \\ & \quad
+2\nabla_\mu\nabla_\nu\phi^I\nabla^\nu\nabla^\lambda\phi^J\nabla_\lambda \nabla^\mu\phi^K
\bigr],\label{eq:multi_generalized_Galileon}
\end{align}
where $G_2$, $G_{3I}$, $G_4$, and $G_{5I}$ are arbitrary functions of
$\phi^I$ and $X^{IJ}:=-g^{\mu\nu}\partial_\mu\phi^I\partial_\nu\phi^J/2$,
and we defined the symmetrized derivative
for any function $f$ of $X^{IJ}$ as
$f_{,\langle IJ\rangle}:=(\partial f/\partial X^{IJ}+\partial f/\partial X^{JI})/2$.
For these functions we must require that
\begin{align}
G_{3I,\langle JK\rangle},\quad G_{4,\langle IJ\rangle,\langle KL\rangle},
\quad
G_{5I,\langle JK\rangle},\quad G_{5I,\langle JK\rangle,\langle LM\rangle}
\end{align}
are symmetric in all of their indices $I,J, \cdots$,
so that the field equations are of second order.
This may be regarded as the multi-scalar version of
the Lagrangian~\eqref{eq:Lag_gen_Galileon},
obtained by generalizing the multi-Galileon theory
on a fixed Minkowski background.

Unlike the case of the single-field Galileon,
the Lagrangian~\eqref{eq:multi_generalized_Galileon}
does {\em not} give the most general multi-scalar-tensor theory
with second-order field equations.
Indeed, the probe-brane derivation of the multi-field
DBI-type Galileon~\cite{RenauxPetel:2011dv,RenauxPetel:2011uk} yields
the terms that cannot be described
by~\eqref{eq:multi_generalized_Galileon} but nevertheless
have second-order field equations~\cite{Kobayashi:2013ina}.
Thus, the ``Galileon route'' is unsuccessful.

In contrast, Ohashi {\em et al.}
followed closely the original derivation of the Horndeski theory
and derived the most general second-order {\em equations of motion}
for {\em bi}-scalar-tensor theory~\cite{Ohashi:2015fma}.
However, the corresponding action has not been obtained so far.

More recently, new terms for the multi-Galileon
that had been overlooked was proposed in~\cite{Allys:2016hfl},
and their covariant completion was obtained in~\cite{Akama:2017jsa}.
These new terms can reproduce the multi-field DBI Galileon~\cite{Akama:2017jsa}.
However, whether or not the most general second-order multi-scalar-tensor
theory is described by the Lagrangian~\eqref{eq:multi_generalized_Galileon}
plus these new terms
remains an open question.

\section{Horndeski theory and cosmology}\label{sec:cosmoH}

A great variety of dark energy/modified gravity models
have been proposed so far to account for the present
accelerated expansion of the universe.
Also in the context of the
accelerated expansion of the early universe, namely, inflation,
gravity modification is now a popular way of
building models. (Actually, one of the earliest proposals of inflation
already invoked higher curvature terms~\cite{Starobinsky:1979ty,Starobinsky:1980te}.)
The Horndeski theory provides us with a useful tool to
study such cosmologies in a unifying way.
In this section, we review the applications of the Horndeski theory
to cosmology.

\subsection{Structure of the background equations}

Let us review the derivation of the field equations for a spatially flat
Friedmann-Lema\^itre-Robertson-Walker (FLRW) metric
and a homogeneous scalar field,
\begin{align}
\D s^2=-N^2(t)\D t^2+a^2(t)\delta_{ij}\D x^i\D x^j,
\quad
\phi = \phi(t).
\end{align}
Here, $N(t)$ can be set to $1$ by redefining the time coordinate,
but we need to retain it for the moment
in order to derive the background equation corresponding to
the Friedmann equation (the $tt$ component of the gravitational
field equations).

First, let us consider the universe filled only with the scalar field.
Substituting this metric and the scalar field to
Eq.~\eqref{eq:Lag_gen_Galileon}, we get the action of the form
\begin{align}
S=\int \D t\D^3x\,L(N,\dot N; a, \dot a, \ddot a; \phi, \dot\phi,\ddot\phi),
\label{eq:action_FLRW}
\end{align}
where a dot denotes differentiation with respect to $t$.
Varying this action with respect to $N$, $a$, and $\phi$,
and then setting $N=1$, we obtain the following
set of the background equations,
\begin{align}
{\cal E}(H;\phi, \dot\phi)&:= -\frac{1}{a^3}\frac{\delta S}{\delta N}=0,
\label{eq:00FLRW}
\\
{\cal P}(H,\dot H; \phi,\dot\phi,\ddot\phi)&:=
\frac{1}{3a^2}\frac{\delta S}{\delta a}=0,
\label{eq:ijFLRW}
\\
{\cal E}_\phi(H,\dot H; \phi,\dot\phi,\ddot\phi)&:=
\frac{1}{Na^3}\frac{\delta S}{\delta \phi} =0,
\label{eq:scalarFLRW}
\end{align}
with $H:=\dot a/a$ being the Hubble parameter.
Equations~\eqref{eq:00FLRW} and~\eqref{eq:ijFLRW}
correspond respectively to the $tt$ and $ij$ components of
the gravitational field equations, and
Eq.~\eqref{eq:scalarFLRW} is the equation of motion for $\phi$.
Explicit expressions for ${\cal E}$, ${\cal P}$, and ${\cal E}_\phi$
are found in~\cite{Kobayashi:2011nu}.

Although the action apparently depends on
$\dot N$, $\ddot a$, and $\ddot\phi$,
all the higher derivatives are canceled in the field equations,
leading to the second-order system as expected.
In particular, ${\cal E}(H;\phi,\dot\phi)=0$ is the constraint equation.
It is interesting to see that
${\cal P}$ and ${\cal E}_\phi$ depend on both $\dot H$ and $\ddot\phi$
in general.
This implies the kinetic mixing of gravity and the scalar field,
which does not occur in Einstein gravity (plus $G_2$).
(In Einstein gravity, ${\cal P}$ (respectively ${\cal E}_\phi$)
is independent of $\ddot\phi$ (respectively $\dot H$).)
In the traditional scalar-tensor theory whose Lagrangian is of the form
${\cal L}=G_2(\phi,X)+G_4(\phi)R$,
this mixing can be undone by moving to the Einstein frame
through the conformal transformation of the metric, $\tilde g_{\mu\nu}=C(\phi)g_{\mu\nu}$.
However, once $G_3(\phi,X)$ is introduced,
the mixing becomes essential and cannot be removed by such
a field redefinition.
This nature
is called {\em kinetic gravity braiding}~\cite{Deffayet:2010qz,Pujolas:2011he}.

Equations~\eqref{eq:00FLRW},~\eqref{eq:ijFLRW}, and~\eqref{eq:scalarFLRW}
are useful for studying the background dynamics of inflation
(and its alternatives), because the early universe can be
described generically by a gravity-scalar system. However,
if one considers the late-time universe based on the Horndeski theory,
it is necessary to introduce the other kind of matter
(dark matter, baryons, and radiation).
In that case, assuming that the matter is minimally coupled to gravity,
the background equations are given by
\begin{align}
{\cal E}=-\rho, \quad {\cal P}=-p,\quad {\cal E}_\phi = 0,
\end{align}
where $\rho$ and $p$ are respectively the energy density and pressure
of the matter.

\subsection{Cosmological perturbations}

\subsubsection{Linear perturbations and stability}

Linear perturbations around a FLRW background are
important in two ways. First, one can judge the stability of
a given cosmological model by studying linear perturbations.
Second, linear perturbations can be used to
test modified theories of gravity against
cosmological observations.
For these purposes let us derive the
quadratic action for linear cosmological perturbations.

Linear perturbations around a FLRW background can be
decomposed into scalar, vector, and tensor components
according to their transformation properties under
three-dimensional spatial rotations (see, e.g.,~\cite{Kodama:1985bj,Mukhanov:1990me}).
The vector perturbations are less interesting because
they are nondynamical
in scalar-tensor theories as well as in Einstein gravity.
We therefore focus on scalar and tensor perturbations.

Thanks to the general covariance, one may make use of the
gauge transformation,
$t\to t-T(t,\Vec{x})$, $\Vec{x}\to\Vec{x}-\Vec{\xi}(t,\Vec{x})$,
to remove some of the perturbation variables. For example,
fluctuations in the scalar field, $\delta\phi(t,\Vec{x})$,
transform as
\begin{align}
\delta\phi\to \delta\phi+\dot\phi T,
\end{align}
and thus we are allowed to take $\delta\phi=0$
by choosing the time coordinate appropriately.
This is called the {\em unitary gauge}, which is particularly useful
and hence we will use for the moment.

Now all the fluctuations are in the metric, and
in the ADM form we parametrize them as
\begin{align}
N=1+\delta N,\quad N_i=\partial_i\psi,\quad
\gamma_{ij}=a^2e^{2\zeta}(e^h)_{ij},
\label{def:metric_pert}
\end{align}
where
\begin{align}
(e^h)_{ij}:=\delta_{ij}+h_{ij}+\frac{1}{2}h_{ik}h_{kj}+\cdots.
\end{align}
Here, $\delta N$, $\psi$, and $\zeta$ are scalar perturbations and
$h_{ij}$ are tensor perturbations (gravitational waves)
satisfying the transverse and traceless conditions,
$\partial^i h_{ij}=0=h_i^i$.
Writing the spatial metric
as $\gamma_{ij}=a^2e^{2\zeta}(e^h)_{ij}$ rather than
$\gamma_{ij}=a^2\left[(1+2\zeta)\delta_{ij}+h_{ij}\right]$
simplifies the computation of the action for the cosmological perturbations.
Note that the spatial gauge transformation was
used to put $\gamma_{ij}$ into the form given above.
Substituting the metric~\eqref{def:metric_pert}
to the action and expanding it to second order in perturbations,
we obtain
\begin{align}
S^{(2)}=S^{(2)}_{\rm tensor} + S^{(2)}_{\rm scalar},
\end{align}
with
\begin{align}
S^{(2)}_{\rm tensor}=
\frac{1}{8}\int\D t\D^3x a^3\left[
{\cal G}_T\dot h_{ij}^2-\frac{{\cal F}_T}{a^2}(\partial_k h_{ij})^2
\right]
\label{action2:tensor}
\end{align}
and
\begin{align}
S^{(2)}_{\rm scalar}=\int \D t\D^3x a^3
&
\biggl[
-3{\cal G}_T\dot\zeta^2+\frac{{\cal F}_T}{a^2}(\partial\zeta)^2
+\Sigma\delta N^2 -2\Theta \delta N \frac{\partial^2\psi}{a^2}
+2{\cal G}_T\dot\zeta\frac{\partial^2\psi}{a^2}
\notag \\ &
+6\Theta\delta N\dot\zeta -2{\cal G}_T\delta N\frac{\partial^2\zeta}{a^2}
\biggr].
\label{action2:scalar1}
\end{align}
The coefficients are given explicitly by
\begin{align}
{\cal G}_T&:=2\left[G_4-2XG_{4X}-X\left(H\dot\phi G_{5X}-G_{5\phi}\right)\right],
\label{stability:gt}
\\
{\cal F}_T&:=2\left[G_4-X\left(\ddot\phi G_{5X}+G_{5\phi}\right)\right],
\label{stability:ft}
\\
\Sigma&:=XG_{2X}+2X^2G_{2XX}+12H\dot\phi XG_{3X}
+6H\dot\phi X^2G_{3XX}
-2XG_{3\phi}-2X^2G_{3\phi X}
\notag \\ & \quad
-6H^2G_4
+6\Bigl[H^2\left(7XG_{4X}+16X^2G_{4XX}+4X^3G_{4XXX}\right)
\notag \\ & \quad
-H\dot\phi\left(G_{4\phi}+5XG_{4\phi X}+2X^2G_{4\phi XX}\right)
\Bigr]
\notag \\ & \quad
+30H^3\dot\phi XG_{5X}+26H^3\dot\phi X^2G_{5XX}+4H^3\dot\phi X^3G_{5XXX}
\notag \\ & \quad
-6H^2X\bigl(6G_{5\phi}
+9XG_{5\phi X}+2 X^2G_{5\phi XX}\bigr),
\\
\Theta&:=-\dot\phi XG_{3X}+
2HG_4-8HXG_{4X}
-8HX^2G_{4XX}+\dot\phi G_{4\phi}+2X\dot\phi G_{4\phi X}
\notag \\ & \quad
-H^2\dot\phi\left(5XG_{5X}+2X^2G_{5XX}\right)
+2HX\left(3G_{5\phi}+2XG_{5\phi X}\right),
\end{align}
which depend on time in general.

We see that
time derivatives of $\delta N$ and $\psi$ do not appear
in the quadratic action for the scalar perturbations~\eqref{action2:scalar1}.
Therefore, variation with respect to $\delta N$ and $\psi$
yields the constraint equations,
\begin{align}
  \Sigma\delta N -\Theta\frac{\partial^2\psi}{a^2}+3\Theta\dot\zeta
  -{\cal G}_T\frac{\partial^2\zeta}{a^2}&=0,
  \\
  \Theta\delta N-{\cal G}_T\dot\zeta&=0.
\end{align}
These equations allows us to express $\delta N$ and $\psi$
in terms of $\zeta$. Then, one can remove $\delta N$ and $\psi$
from Eq.~\eqref{action2:scalar1} and obtain the action
written solely in terms of $\zeta$ (the curvature perturbation
in the unitary gauge):
\begin{align}
S_\zeta^{(2)}=
\int \D t\D^3x a^3 \left[
{\cal G}_S\dot\zeta^2-\frac{{\cal F}_S}{a^2}(\partial\zeta)^2
\right],
\label{action2:scalar2}
\end{align}
where
\begin{align}
{\cal G}_S&:=\frac{\Sigma}{\Theta^2}{\cal G}_T^2+3{\cal G}_T,
\label{stability:gs}
\\
{\cal F}_S&:=\frac{1}{a}\frac{\D}{\D t}\left(\frac{a}{\Theta}{\cal G}_T^2\right)-{\cal F}_T.
\label{stability:fs}
\end{align}
The general quadratic actions~\eqref{action2:tensor}
and~\eqref{action2:scalar2} were
derived in~\cite{Kobayashi:2011nu}.

It is instructive to check here that the standard textbook result
is reproduced in the case of general relativity $+$
a canonical scalar field, $G_2=X-V(\phi)$, $G_4=\mpl^2/2$, $G_3=G_5=0$.
Obviously, we have ${\cal G}_T={\cal F}_T=\mpl^2$.
Since $\Sigma = X-3\mpl^2H^2$ and $\Theta = \mpl^2H$,
we have ${\cal G}_S=X/H^2$ and ${\cal F}_S=\mpl^2(-\dot H/H^2)=\mpl^2\epsilon$,
where $\epsilon:=-\dot H/H^2$ is the slow-roll parameter.
Using the background equation, $-\mpl^2\dot H=X$,
it turns out that ${\cal G}_S={\cal F}_S=\mpl^2\epsilon$,
and thus the standard result is obtained.

The propagation speeds of the tensor and scalar modes are
given respectively by
\begin{align}
c_{\rm GW}^2&:=\frac{{\cal F}_T}{{\cal G}_T},\label{speedofGW_H}
\\
c_s^2&:=\frac{{\cal F}_S}{{\cal G}_S}.
\end{align}
These quantities must be positive,
$c_{\rm GW}^2>0$, $c_s^2>0$, because otherwise
each perturbation mode exhibits an exponential growth.
This is called the gradient instability.
This instability is dangerous in particular for
short-wavelengh modes, because
the time scale of the instability is proportional to the wavelength.

In addition to the above stability conditions,
we require that ${\cal G}_T>0$ and ${\cal G}_S>0$
in order to guarantee the positivity of the kinetic terms for
$h_{ij}$ and $\zeta$, i.e., the absence of
ghost instabilities.
To sum up, for a given cosmological model to be stable,
one must demand that
\begin{align}
{\cal G}_T>0,\quad {\cal F}_T>0,
\quad {\cal G}_S>0,\quad {\cal F}_S>0.\label{stablity_conditions}
\end{align}

The equation of motion derived from~\eqref{action2:scalar2} is
\begin{align}
\frac{1}{a^3{\cal G}_S}\frac{\D}{\D t}\left(
a^3{\cal G}_S\frac{\D\zeta}{\D t}
\right)-\frac{c_s^2}{a^2}\partial^2\zeta=0.
\end{align}
On large (superhorizon\footnote{Since the sound speed $c_s$ is
different from 1 in general, the horizon scale here should be understood as
the sound horizon scale. The same remark applies to the tensor modes.})
scales, one may ignore the second term and
obtain the solution
\begin{align}
\zeta(t,\Vec{x})\simeq C(\Vec{x})+D(\Vec{x})\int^t \frac{\D t'}{a^3(t'){\cal G}_S(t')},
\label{sol:superhorizon_zeta}
\end{align}
where $C$ and $D$ are integration functions.
It is natural to assume that
all the time-dependent functions
(except, of course, for the scale factor, $a\sim e^{Ht}$)
vary slowly during inflation,
and hence ${\cal G}_S\simeq\,$const.
If this is the case, the second term in Eq.~\eqref{sol:superhorizon_zeta}
decays rapidly and thus can be neglected,
leading to the conservation of $\zeta$ on superhorizon scales,
$\dot\zeta\simeq 0$, in the Horndeski theory~\cite{Naruko:2011zk,Gao:2011mz}.
This is the generalization of the standard result.
Note, however, that even in the case of
general relativity + a canonical scalar field
the ultra slow-roll/nonattractor phase of inflation can appear,
in which we have ${\cal G}_S\propto a^{-6}$ and
the second term grows~\cite{Inoue:2001zt,Kinney:2005vj}.
The nonattractor inflationary dynamics may be more
complicated in the presence of the Galileon terms~\cite{Hirano:2016gmv}.

Similarly, for tensor perturbations we have the
superhorizon solution,
\begin{align}
h_{ij}(t,\Vec{x})\simeq
C_{ij}(\Vec{x})+D_{ij}(\Vec{x})\int^t \frac{\D t'}{a^3(t'){\cal G}_T(t')},
\end{align}
where $C_{ij}$ and $D_{ij}$ are integration functions,
and we see that the second term corresponds to the decaying mode.
However, ${\cal G}_T$ can vary rapidly in time
in some ultra slow-roll models of inflation with
nonminimal couplings between gravity and the scalar field
(i.e., nonconstant $G_4$ and $G_5$).
In such a model,
the would-be decaying tensor mode can grow
in a similar manner to the aforementioned growth of $\zeta$~\cite{Mylova:2018yap}.

For the purpose of computing the power spectra of
primordial perturbations from inflation,
it is convenient to recast the quadratic
actions~\eqref{action2:tensor} and~\eqref{action2:scalar2}
into the canonically normalized form.
For $\zeta$ we introduce
the new time coordinate defined by $\D y:=(c_s/a)\D t$ and
variable
\begin{align}
u:=z\zeta,\quad z:=\sqrt{2}a({\cal F}_S{\cal G}_S)^{1/4}.
\end{align}
Then, we have
\begin{align}
S_\zeta^{(2)}=\frac{1}{2}\int \D y \D^3x\left[
(u')^2-(\partial u)^2+\frac{z''}{z}u^2
\right],
\end{align}
where a prime here denotes differentiation with respect to $y$.
This is of the familiar ``Sasaki-Mukhanov'' form.
Tensor perturbations can be analyzed in a similar way~\cite{Kobayashi:2011nu}.

The power spectrum can be evaluated
by following the standard procedure
to quantize $u$~\cite{Mukhanov:1990me}.
Let us assume for simplicity that
the time dependent coefficients in the quadratic action
vary very slowly during inflation.
In such a ``slow-varying'' limit, the power spectra for
the curvature and tensor perturbations are given respectively by
\begin{align}
{\cal P}_\zeta = \frac{{\cal G}_S^{1/2}}{2{\cal F}_S^{3/2}}\frac{H^2}{4\pi^2},
\\
{\cal P}_h = \frac{8{\cal G}_T^{1/2}}{{\cal F}_T^{3/2}}\frac{H^2}{4\pi^2},
\end{align}
evaluated at the (sound) horizon crossing time.
The tensor-to-scalar ratio $r:={\cal P}_h/{\cal P}_\zeta$ is given by
\begin{align}
r=16\left(\frac{{\cal F}_S}{{\cal F}_T}\right)^{3/2}
\left(\frac{{\cal G}_S}{{\cal G}_T}\right)^{-1/2}.
\end{align}
The standard expression $r=16\epsilon$ can be reproduced by substituting
${\cal G}_T={\cal F}_T=\mpl^2$ and ${\cal G}_S={\cal F}_S=\mpl^2\epsilon$,
but in general the tensor-to-scalar ratio and the consistency relation
can be nonstandard in the Horndeski theory.

\subsubsection{Beyond linear order}

With increasingly precise measurements of CMB anisotropy,
it is important to study non-Gaussian signatures of
primordial perturbations from inflation.
For this purpose we need to compute the action
to cubic (and higher) order in perturbations.
Following the seminal work by Maldacena~\cite{Maldacena:2002vr},
this program can be carried out in the context of the Horndeski theory.

The cubic action for the scalar perturbations in the Horndeski theory is
given in~\cite{Gao:2011qe,DeFelice:2011uc}.
It was pointed out in~\cite{RenauxPetel:2011sb} that no new operators appear
compared to simpler k-inflation~\cite{Seery:2005wm,Chen:2006nt},
though we have four free functions in the theory so that
there is a larger degree of freedom in adjusting
the coefficients of each term in the cubic action.
Shapes of non-Gaussianities have been investigated in more detail
in~\cite{Ribeiro:2011ax,DeFelice:2013ar}.

The cubic action for the tensor perturbations
is presented in~\cite{Gao:2011vs}, where it was found that
only two independent operators appear including the one
that is already present in general relativity.
The non-Gaussian contribution from the new term in the Horndeski theory
might in principle be detectable through CMB B-mode polarization
if the corresponding coefficient is extremely large~\cite{Tahara:2017wud}.

Cross-bispectra among tensor and scalar perturbations
were computed in~\cite{Gao:2012ib}.

\subsection{NEC-violating cosmologies and their stability}\label{sec:nogotheo}

In this subsection, we discuss cosmological consequences of
violation of the null energy condition (NEC) based on
the Horndeski theory. See also Ref.~\cite{Rubakov:2014jja}
for a mini-review on the same subject.

The NEC demands the following
bound on the energy momentum tensor:
\begin{align}
T_{\mu\nu}k^\mu k^\nu \ge 0
\end{align}
for any null vector $k^\mu$.
In the context of cosmology, this condition is equivalent to
\begin{align}
\rho + p \ge 0,
\end{align}
and then in general relativity the NEC implies that
\begin{align}
\dot H\le 0\label{NEC-H}
\end{align}
through the Einstein equations.
As there is no clear distinction between
the energy-momentum tensor of the scalar field and the ``left-hand side''
(i.e., the geometrical part)
of the gravitational field equations in scalar-tensor theories,
in the following we mean Eq.~\eqref{NEC-H} by the NEC.

In usual inflationary cosmology with a canonical scalar field,
we have $-2\mpl^2\dot H = \rho+p=\dot\phi^2\ge 0$,
and thus the NEC is automatically satisfied.
This in particular implies that the spectrum of primordial tensor modes,
${\cal P}_h=2H^2/\pi^2\mpl^2$ evaluated at horizon crossing,
must be red.
If the spectrum of the tensor modes would have a blue tilt,
the NEC might be violated during inflation due to some
nonstandard mechanism.

Although inflation is a very attractive scenario,
inflationary spacetime is past incomplete~\cite{Borde:1993xh,Borde:1996pt,Borde:2001nh}
(see also~\cite{Yoshida:2018ndv}),
which motivates
nonsingular alternatives to inflation such as bouncing
models (see, e.g.,~\cite{Lehners:2008vx,Novello:2008ra,%
Brandenberger:2009jq,Battefeld:2014uga,Cai:2014bea} for a review).
In nonsingular cosmologies, there must be some interval
during which the NEC is violated.
A noncanonical scalar field or some other kind of matter
is required to realize such nonsingular alternatives.

It is therefore interesting to explore the possibilities of
NEC-violating cosmology in scalar-tensor theories.
Since we would expect that the energy conditions are somehow related to
the stability of spacetime, now
the key question is: can we construct {\em stable}
NEC-violating cosmology?

To answer this question, let us first consider
Einstein gravity plus $G_2(\phi, X)$.
The background equations in this case read
\begin{align}
3\mpl^2H^2&=2XG_{2X}-G_2 ,
\\
-\mpl^2\left(3H^2+2\dot H\right) &= G_2 ,
\end{align}
and hence $-\mpl^2 \dot H = XG_{2X}$.
The NEC can therefore be violated if the function $G_2(\phi,X)$ is chosen so that
$G_{2X}<0$ can occur. However, in this theory we have
${\cal F}_S=\mpl^2(-\dot H/H^2)$,
which implies that NEC-violating solutions are unstable.
Thus, in this simplest case the NEC is closely related to stability.

The situation drastically changes if one adds
$G_3$ and the other more general terms that are included in the Horndeski theory,
because the general expressions for the stability conditions,
\eqref{stability:gt},~\eqref{stability:ft},~\eqref{stability:gs}
and~\eqref{stability:fs}, are not correlated with the sign of $\dot H$.
Therefore, one can construct NEC-violating
stages that are nevertheless stable within the Galileon and Horndeski
theories~\cite{Nicolis:2009qm} (see, however,~\cite{Sawicki:2012pz}).
This opens up Pandora's box of
nonsingular bouncing cosmology~\cite{Qiu:2011cy,Easson:2011zy,Osipov:2013ssa}
as well as blue gravitational waves from inflation~\cite{Kobayashi:2010cm}
(see, however,~\cite{Cai:2014uka}).

A novel NEC-violating cosmological scenario
called the {\it galilean genesis} was proposed based on the cubic Galileon
theory~\cite{Creminelli:2010ba}.
The Lagrangian for this scenario is given by
\begin{align}
{\cal L}=\frac{\mpl^2}{2}R-e^{2\phi/f}X-\frac{X}{\Lambda^3}\Box\phi+
\frac{X^2}{\Lambda^3f},
\end{align}
where $f$ and $\Lambda$ are positive constants having the dimension of mass.
This theory admits the following approximate solution
valid for $\mpl(-t)\gg (f/\Lambda)^{3/2}$:
\begin{align}
a\simeq 1 + \frac{f^3}{8\mpl^2\Lambda^3}\frac{1}{(-t)^2},
\quad
H\simeq \frac{f^3}{4\mpl^2\Lambda^3}\frac{1}{(-t)^3},
\quad
e^{\phi/f}\simeq \sqrt{\frac{3f}{2\Lambda^3}}\frac{1}{(-t)}
\quad (t<0).\label{genesis_solution}
\end{align}
As seen from~\eqref{genesis_solution}, the universe starts expanding from
a low energy, quasi-Minkowski state in the asymptotic past.
Clearly, the NEC is violated. Nevertheless, we have
\begin{align}
{\cal G}_S\simeq {\cal F}_S\simeq 12\mpl^4(-t)^2\left(\frac{\Lambda}{f}\right)^3>0,
\end{align}
showing that this solution is stable.
The galilean genesis thus has a potential to be an interesting alternative
to inflation.
This scenario has further been generalized
and investigated in more detail in
Refs.~\cite{LevasseurPerreault:2011mw,Liu:2011ns,Hinterbichler:2012mv,%
Wang:2012bq,Liu:2012ww,Creminelli:2012my,Hinterbichler:2012fr,%
Hinterbichler:2012yn,Liu:2013xt,Easson:2013bda,Nishi:2014bsa,%
Nishi:2015pta,Nishi:2016wty,Nishi:2016ljg,Ageeva:2018lko}.

 As an alternative to inflation,
the genesis phase described by~\eqref{genesis_solution}
is supposed to be matched onto a radiation-dominated universe
across the reheating stage at $t\sim -(f/\Lambda)^{3/2}/\mpl$.
Or, one may consider the initial genesis phase
followed by inflation as an ``early-time completion''
of the inflationary scenario~\cite{Pirtskhalava:2014esa}.
In any case, a problem arises in considering the
whole history of such a singularity-free universe:
gradient instabilities show up at some moment in the history.
Not only the galilean genesis but also bouncing models
have the same problem. Several examples show that
instabilities may occur at the transition
from the NEC-violating phase to some subsequent phase
or even in a far future after the transition~\cite{Easson:2011zy,Pirtskhalava:2014esa,%
Cai:2012va,Koehn:2013upa,Battarra:2014tga,Qiu:2015nha,Wan:2015hya,Kobayashi:2015gga}.
This implies that the instabilities may not be related directly to
the violation of the energy condition.

In fact,
it can be proven that the appearance of gradient instabilities
is generic to all nonsingular cosmological solutions in
the Horndeski theory~\cite{Libanov:2016kfc,Kobayashi:2016xpl}
(see also~\cite{Rubakov:2013kaa,Elder:2013gya}).
As we have seen, one can construct a NEC-violating solution
that is stable during a finite interval. What we will observe below is that
such a solution is however unstable once the whole history is concerned.
The key inequality follows from Eq.~\eqref{stability:fs} and
the stability conditions:
\begin{align}
\frac{\D\xi}{\D t}>a{\cal F}_T>0
\quad (-\infty < t<\infty)
,\label{ineq:nons}
\end{align}
where $\xi:=a{\cal G}_T^2/\Theta$.
For a stable, nonsingular cosmological solution we have
$a\ge \,$const, ${\cal G}_T>0$, and $|\Theta|<\infty$.
Therefore, $\xi$ must be a monotonically increasing
function of time that never crosses zero,
which means that $\xi\to\,$const as $t\to\infty$ or $-\infty$.
(Note that $\Theta$ and hence $\xi$ can take either sign.)
Integrating Eq.~\eqref{ineq:nons}
from $-\infty$ to some $t$ and from some $t$ to $\infty$,
one obtains
\begin{align}
\xi(t)-\xi(-\infty)>\int^t_{-\infty}a{\cal F}_T\D t,
\quad
\xi(\infty)-\xi(t)>\int^{\infty}_ta{\cal F}_T\D t.
\label{ineq:geogw}
\end{align}
At least either of the integrals must be convergent.
Otherwise the stability conditions would be violated
at some moment in the entire history of the universe.

By designing the functions in the Horndeski action
so that either of the integrals in~\eqref{ineq:geogw}
is convergent, it is indeed possible to construct a stable, nonsingular cosmological
solution~\cite{Kobayashi:2016xpl,Ijjas:2016vtq}.
However, the convergent integral indicates that
the spacetime is geodesically incomplete for the propagation of
gravitons~\cite{Creminelli:2016zwa}.
This can be understood by moving to the Einstein frame for
gravitons via disformal transformation~\cite{Creminelli:2014wna}.
Moreover,
the normalization of vacuum quantum fluctuations tells us that they would grow and diverge
if ${\cal F}_T$ approaches zero sufficiently fast
either in the asymptotic past or the future, which implies that
the tensor sector is pathological.
If one requires geodesic completeness for gravitons
and thereby avoids this subtle behavior, then
stable, nonsingular cosmologies are prohibited within the Horndeski theory.
Note that this no-go theorem cannot tell when the gradient instability
shows up. That moment may be in the remote future from
the early NEC-violating phase.

Several comments are now in order. First,
the no-go theorem for nonsingular cosmologies can be extended to
include multiple components other than the Horndeski
scalar~\cite{Creminelli:2016zwa,Kolevatov:2016ppi,Akama:2017jsa} and to
the spatially open universe~\cite{Akama:2018cqv}.
Second, there is some debate about zero-crossing of $\Theta$
and the validity of the use of the curvature perturbation
in the unitary gauge $\zeta$~\cite{Battarra:2014tga,Quintin:2015rta,%
Ijjas:2017pei,Dobre:2017pnt,Mironov:2018oec}.
Third, from the effective field theory viewpoint,
the strong coupling scale may cut off the instabilities~\cite{Koehn:2015vvy}
(see also~\cite{deRham:2017aoj}).
Finally, the no-go theorem can be circumvented in
scalar-tensor theories beyond Horndeski~\cite{Creminelli:2016zwa,Cai:2016thi,Cai:2017tku,%
Cai:2017dyi,Kolevatov:2017voe,Mironov:2018oec,Ye:2019frg}.\footnote{As will be argued
in Sec.~\ref{sec:dhost}, only theories that can be generated from
the Horndeski theory via disformal transformation~\eqref{def:disformalTr}
are phenomenologically viable.
Since the disformal transformation is just a field redefinition,
one may wonder why the the no-go theorem can be evaded in theories beyond
(and disformally related to) Horndeski. The trick is that the disformal
transformation that generates the theories admitting stable nonsingular cosmology
is singular at some moment~\cite{Creminelli:2016zwa}.}

\subsection{Inclusion of matter}

So far we have considered cosmological perturbations
in the universe dominated by $\phi$,
bearing the application to the early universe in mind.
Let us extend the previous results to include
the other kind of matter,
since in the late universe matter perturbations would be also important.

\subsubsection{Stability conditions}

As additional matter, we are interested in an irrotational, barotropic
perfect fluid minimally coupled to the metric.
Such a fluid can be mimicked by a k-essence field $\chi$
whose action $S_m$ is of the form
\begin{align}
S_m=\int \D^4x\sqrt{-g}P(Y), \quad Y:=-\frac{1}{2}g^{\mu\nu}\partial_\mu\chi\partial_\nu\chi.
\end{align}
Introducing the k-essence field as a perfect fluid
is a concise and useful technique to treat
the fluid at the action level.
The energy-momentum tensor of $\chi$
is given by $T_{\mu\nu}=2P_Y\partial_\mu\chi \partial_\nu\chi +Pg_{\mu\nu}$,
from which we see that
the energy density, pressure, and four-velocity of this fluid are
expressed as
$\rho=2YP_Y-P$, $p=P$, and $u_\mu=-\partial_\mu\chi/\sqrt{2Y}$.
The background equation for $\chi$,
which is equivalent to $\nabla_\nu T^\nu_{\mu}=0$, reads
\begin{align}
  \frac{\D}{\D t}\left(a^3P_Y\dot\chi\right)=0\quad \Rightarrow\quad
\ddot\chi+3c_m^2H\dot\chi = 0,\label{eom_background_chi}
\end{align}
where
\begin{align}
c_m^2:=\frac{\dot p}{\dot \rho}=\frac{P_Y}{P_Y+2YP_{YY}},
\end{align}
is the sound speed squared of the matter.
For $P\propto Y^n$, we have
$w:=p/\rho=\,$const$\,=1/(2n-1) (=c_m^2)$~\cite{Matarrese:1984zw,Garriga:1999vw}.
This implies that one must be careful when taking the limit of
pressureless dust, $c_m^2,w\to 0$, which is singular (see~\cite{Boubekeur:2008kn}).
We therefore assume that $c_m^2\neq 0$ for the moment.

Expanding the action to second order in perturbations,
we obtain $S^{(2)}=S_{\rm tensor}^{(2)}+S_{\rm scalar}^{(2)}+S_m^{(2)}$,
where $S_{\rm tensor}^{(2)}$ and $S_{\rm scalar}^{(2)}$
are given by Eqs.~\eqref{action2:tensor} and~\eqref{action2:scalar1},
respectively. The contribution from the matter action, $S_m^{(2)}$, is given by
\begin{align}
S_m^{(2)}&=\int \D t\D^3x
\frac{a^3P_Y}{c_m^2}\left[
Y\delta N^2-\dot\chi\left(\delta N-3c_m^2\zeta\right)\dot{\delta\chi}
+c_m^2\dot\chi \frac{\partial^2\psi}{a^2}\delta\chi+\frac{1}{2}
\dot{\delta\chi}^2-\frac{c_m^2}{2a^2}(\partial\delta\chi)^2
\right],\label{action:chi2}
\end{align}
where $\delta\chi=\delta\chi(t,\Vec{x})$ is a fluctuation of $\chi$.
Since the tensor sector remains unaltered by the inclusion of the matter,
we focus on the scalar sector.

It follows from $\delta S^{(2)}/\delta(\delta N)=0$
and $\delta S^{(2)}/\delta\psi = 0$ that
\begin{align}
  \Sigma\delta N -\Theta\frac{\partial^2\psi}{a^2}+3\Theta\dot\zeta
  -{\cal G}_T\frac{\partial^2\zeta}{a^2}+\frac{YP_Y}{c_m^2}
\delta N  -\frac{P_Y}{2c_m^2}\dot\chi\dot{\delta\chi}&=0,
  \\
  \Theta\delta N-{\cal G}_T\dot\zeta
-\frac{P_Y}{2}\dot\chi{\delta\chi}
  &=0.\label{momentum_const_matter}
\end{align}
Similarly to the previous analysis without $\chi$,
one can eliminate
$\delta N$ and $\psi$ from the quadratic action
by using Eq.~\eqref{momentum_const_matter}.
The reduced action written solely in terms of $\zeta$ and $\delta\chi$
takes the form~\cite{DeFelice:2011bh}
\begin{align}
S^{(2)}
&=\int\D t\D^3x a^3 \left[
G_{AB}\dot q^A\dot q^B-\frac{1}{a^2}F_{AB}\partial q^A\cdot\partial q^B+\cdots
\right],
\end{align}
where
\begin{align}
q^A:=\left(\zeta, \frac{{\cal G}_T}{\Theta}\frac{\delta\chi}{\dot\chi} \right),
\end{align}
and
\begin{eqnarray}
  G_{AB} = \left(
    \begin{array}{cc}
      {\cal G}_S+Z & -Z \\
      -Z & Z
    \end{array}
  \right),
\quad
  F_{AB} = \left(
    \begin{array}{cc}
      {\cal F}_S & -c_m^2Z \\
      -c_m^2Z & c_m^2Z
    \end{array}
  \right),
\end{eqnarray}
with
\begin{align}
Z:=\left(\frac{{\cal G}_T}{\Theta}\right)^2\frac{\rho+p}{2c_m^2}.
\end{align}
Here we only write the terms that are relevant to ghost and gradient instabilities.
To avoid ghost instabilities we require that $G_{AB}$ is a
positive definite matrix. This is equivalent to
${\cal G}_S>0$ and $Z>0$.
The propagation speeds of the two scalar modes are determined by solving
det$(v^2G_{AB}-F_{AB})=0$, yielding
$v^2=({\cal F}_S-c_m^2Z)/{\cal G}_S$ and $v^2=c_m^2$.
Thus, the stability conditions in the presence of
an additional perfect fluid are summarized as
\begin{align}
{\cal G}_S>0,\quad \rho+p>0,\quad c_m^2>0,
\quad {\cal F}_{S}> \frac{1}{2}\left(\frac{{\cal G}_T}{\Theta}\right)^2(\rho+p).
\end{align}
It can be seen that the conditions imposed on the fluid component are quite reasonable.

\subsubsection{Matter density perturbations}

In late-time cosmology, we are often interested in the evolution
of the density perturbations of pressureless matter on subhorizon scales.
The analysis is usually done in the Newtonian gauge,
in which the metric takes the form
\begin{align}
\D s^2=-[1+2\Phi(t,\Vec{x})]\D t^2+a^2[1-2\Psi(t,\Vec{x})]\delta_{ij}\D x^i\D x^j,
\label{metric:Newtonian}
\end{align}
with the nonvanishing scalar-field fluctuation,
\begin{align}
\phi=\phi(t)+\delta\phi(t,\Vec{x}).
\end{align}
One can move from the unitary gauge to the Newtonian gauge
by performing the coordinate transformation
$t_N=t-T$ such that
\begin{align}
\Phi=\delta N+\dot T,\quad \Psi=-\zeta-HT,
\quad 0=\psi-T,\quad \delta\phi=0+\dot\phi T.\label{eq:uni-to-N1}
\end{align}
The fluctuation of $\chi$ in the Newtonian gauge is given by
\begin{align}
\delta \chi_N=\delta \chi+\dot\chi T.\label{eq:uni-to-N2}
\end{align}

Substituting Eqs.~\eqref{eq:uni-to-N1} and~\eqref{eq:uni-to-N2}
to Eqs.~\eqref{action2:scalar1} and~\eqref{action:chi2},
we obtain the Newtonian gauge expression for the quadratic action.
As we are interested in the quasi-static evolution of the perturbations
inside the (sound) horizon, we assume that
$\dot\varepsilon\sim H\varepsilon \ll \partial\varepsilon/a$
($\varepsilon=\Phi,\Psi, H\delta\phi/\dot\phi$).
We will take the pressureless limit $c_m^2\to0$,
which is apparently singular. Therefore, we retain carefully the
would-be singular terms in this limit.
The resultant action in the quasi-static approximation is given by
\begin{align}
S^{(2)}_{\rm QS}=
\int\D t\D^3x&\biggl\{a \left[
{\cal F}_T(\partial\Psi)^2-2{\cal G}_T\partial\Phi\partial\Psi
+b_0H^2(\partial T)^2
-2b_1H\partial T\partial\Psi
-2b_2H\partial T\partial\Phi
\right]
\notag \\ &  +
\frac{a^3 P_Y}{2} \left[
 -\frac{1}{a^2}(\partial\delta\chi_N)^2 +
\frac{1}{c_m^2}
 \left(\dot{\delta\chi}_N-\dot\chi\Phi\right)^2
\right]\biggr\},\label{action:matter_density}
\end{align}
where $T=\delta\phi/\dot\phi$ and the coefficients are
defined as
\begin{align}
b_0&:=\frac{1}{H^2}\left[
\dot\Theta+H\Theta +H^2({\cal F}_T-2{\cal G}_T)-2H\dot{\cal G}_T
+YP_Y
\right],\label{def:b0}
\\
b_1&:=\frac{1}{H}\left[\dot{\cal G}_T+H({\cal G}_T-{\cal F}_T)\right],
\\
b_2&:=\frac{1}{H}\left(H{\cal G}_T-\Theta \right).
\end{align}
In Eq.~\eqref{def:b0} one may replace $YP_Y$ with $\rho/2$
in the pressureless limit.
Note that there could be terms
of the form $m^2\varepsilon^2$ (without spatial derivatives)
which are larger than ${\cal O}(H^2\varepsilon^2)$
and can be as large as ${\cal O}((\partial\varepsilon)^2)$,
but for simplicity we ignored such terms.

The equations of motion derived from~\eqref{action:matter_density} are
\begin{align}
  \delta\Psi:&\quad
\partial^2\left({\cal F}_T\Psi-{\cal G}_T\Phi -b_1H T\right)=0,
\label{eom1_density_pert}
\\
\delta \Phi:&\quad
\partial^2\left({\cal G}_T\Psi +b_2HT\right)=
\frac{a^2P_Y}{2c_m^2}\left(\dot\chi\dot{\delta\chi}_N-\dot\chi^2\Phi\right)
\;\left(=\frac{\delta\rho}{2}\right),
\label{eom2_density_pert}
\\
\delta T:&\quad
\partial^2\left(b_0HT-b_1\Psi-b_2\Phi \right)=0,
\label{eom3_density_pert}
\end{align}
and
\begin{align}
\delta \chi_N:\quad
\frac{\D}{\D t}\left[\frac{a^3P_Y}{c_m^2}\left(\dot{\delta\chi}_N-\dot\chi\Phi\right)
\right]=aP_Y\partial^2\delta\chi_N,\label{eom4_density_pert}
\end{align}
where the right-hand side of Eq.~\eqref{eom2_density_pert}
may be replaced with the density perturbation by noting that
\begin{align}
\delta\rho =\frac{P_Y}{c_m^2}\left(\dot\chi\dot{\delta\chi}_N-\dot\chi^2\Phi\right).
\end{align}
Thus, the apparently singular behavior pf this equation
in the $c_m^2\to 0$ limit
can be eliminated.
Taking $b_0=b_1=b_2=0$
in Eqs.~\eqref{eom1_density_pert}--\eqref{eom3_density_pert},
the standard result in Einstein gravity
can be recovered.

Solving the algebraic equations~\eqref{eom1_density_pert}--\eqref{eom3_density_pert},
one arrives at the modified Poisson equation~\cite{DeFelice:2011hq},
\begin{align}
\frac{1}{a^2}\partial^2\Phi = 4\pi G_{\rm eff}(t)\delta\rho,
\quad
8\pi G_{\rm eff}:=\frac{b_0{\cal F}_T-b_1^2}{b_0{\cal G}_T^2+2b_1b_2{\cal G}_T+b_2^2{\cal F}_T}.
\label{eq:mod_Poiss}
\end{align}
The effective gravitational coupling
$G_{\rm eff}$ can be different from the Newton constant,
and, as seen below, it affects the evolution of the density perturbations.
The ratio
\begin{align}
\eta(t):=\frac{\Psi}{\Phi }
=\frac{b_0{\cal G}_T+b_1b_2}{b_0{\cal F}_T-b_1^2}
\end{align}
is also an important quantity because
$\eta\neq 1$ if gravity is nonstandard.
Since the bending of light depends on $\Phi+\Psi$,
weak-lensing observations are useful to test $\eta\neq 1$.
Note that the above expressions cannot be used
for $f(R)$ and chameleon models of dark energy, because
we dropped the mass term for simplicity.
See Refs.~\cite{Noller:2013wca,Sawicki:2015zya,Chiu:2015voa}
for the limits of the quasi-static assumption
and Refs.~\cite{DeFelice:2011hq,Gleyzes:2013ooa,Bellini:2014fua}
for the complete expression of the equations (without
using the quasi-static approximation).


The equation of motion~\eqref{eom4_density_pert} can be written as
\begin{align}
\frac{\D}{\D t}\left(a^3\delta\rho\right)
-3c_m^2H\left(a^3\delta\rho\right)
= aP_Y\dot\chi \partial^2\delta\chi_N,
\end{align}
where we used~\eqref{eom_background_chi}.
Using~\eqref{eom_background_chi} again, one finds
\begin{align}
\frac{\D}{\D t}\left\{a^2\left[
\frac{\D}{\D t}\left(a^3\delta\rho\right)
-3c_m^2H\left(a^3\delta\rho\right)
\right]\right\}=a^3 \partial^2\left(c_m^2\delta\rho + 2YP_Y\Phi\right).
\end{align}
Now we can take the limit $c_m^2\to 0$ and $2YP_Y\to\rho$ safely to get
\begin{align}
\ddot\delta + 2H\dot \delta =  \frac{1}{a^2}\partial^2\Phi,
\label{eq:evol_dens}
\end{align}
where $\delta:=\delta\rho/\rho$.
Thus, as expected, the familiar evolution equation for
the density contrast $\delta$ is recovered from the
equation of motion for $\delta\chi_N$ in the pressureless limit.
Combining this with the modified Poisson equation~\eqref{eq:mod_Poiss},
one can derive the closed-form evolution equation for $\delta$.

Instead of introducing the k-essence field $\chi$,
one may replace
the second line in Eq.~\eqref{action:matter_density}
with
\begin{align}
-a^3\Phi\delta\rho,
\end{align}
and use the usual fluid equations for a pressureless dust.
This is a simpler procedure to arrive at the same result.

\section{Beyond Horndeski}\label{sec:beyondH}

So far we have considered the most general scalar-tensor theory
having second-order equations of motion and its application to cosmology.
Thanks to this second-order nature, the theory is obviously free of
the Ostrogradsky instability.
However, it should be emphasized that
the second-order equations of motion
are {\em not} the necessary conditions for the absence of
the Ostrogradsky instability in theories with multiple fields.

To see this, let us consider a simple toy model
in mechanics whose Lagrangian is given by~\cite{Langlois:2015cwa}
\begin{align}
L=\frac{a}{2}\ddot\phi^2+b\ddot\phi \dot q+\frac{c}{2}\dot q^2+\frac{1}{2}\dot\phi^2
-\frac{1}{2}\phi^2-\frac{1}{2}q^2.
\end{align}
Here, the coefficients $a$, $b$, and $c$ are assumed to be constants.
The Euler-Lagrange equations are of higher order in general:
\begin{align}
a\ddddot\phi +b\dddot q - \ddot\phi  - \phi &=0,\label{deg-eom1}
\\
b\dddot \phi + c\ddot q + q&=0,\label{deg-eom2}
\end{align}
implying that the system contains an extra degree of freedom
and hence suffers from the Ostrogradsky ghost.
However, if the kinetic matrix constructed from the highest derivative terms,
\begin{align}
M=
\begin{pmatrix}
  a & b \\ b & c
\end{pmatrix},
\end{align}
is degenerate, i.e., $ac-b^2=0$, then the system
contains only 2 degrees of freedom.
Indeed, if $ac-b^2=0$ is satisfied, we can combine the equations of
motion~\eqref{deg-eom1} and~\eqref{deg-eom2} to reduce the number of
derivatives. First, $c\times$\eqref{deg-eom1}$-b\times \D$\eqref{deg-eom2}$/\D t$
gives
\begin{align}
\ddot\phi +\frac{b}{c}\dot q +   \phi=0.\label{deg-eom3}
\end{align}
Then, $\D$\eqref{deg-eom3}$/\D t$ is used to remove $\dddot\phi$
from Eq.~\eqref{deg-eom2}, yielding
\begin{align}
  \left(1-\frac{b^2}{c^2}\right)\ddot q -\frac{b}{c}\dot\phi + \frac{1}{c}q=0.
  \label{deg-eom4}
\end{align}
We thus arrive at the two second-order equations of
motion~\eqref{deg-eom3} and~\eqref{deg-eom4}
for $\phi$ and $q$.
This shows that the degenerate system is free of the Ostrogradsky ghost
and hence is healthy despite the higher-order Euler-Lagrange equations.

In this section, we will briefly
explore such healthy degenerate higher-order theories
containing the metric and a scalar field
and extend the Horndeski theory.
The reader is referred to~\cite{Langlois:2017mdk,Langlois:2018dxi} for a more complete review
on this topic.

\subsection{Degenerate higher-order scalar-tensor theories}\label{sec:dhost}

The first example of degenerate higher-order
scalar-tensor (DHOST) theories beyond Horndeski~\cite{Zumalacarregui:2013pma}
was obtained by performing a disformal transformation~\cite{Bekenstein:1992pj}
\begin{align}
g_{\mu\nu}\to \tilde g_{\mu\nu}=C(\phi,X)g_{\mu\nu}+D(\phi,X)\phi_\mu\phi_\nu.
\label{def:disformalTr}
\end{align}
This is a generalization of the familiar conformal transformation,
$\tilde g_{\mu\nu}=C(\phi)g_{\mu\nu}$.
The disformal transformation~\eqref{def:disformalTr}
is invertible if
\begin{align}
C(C-XC_X+2X^2D_X)\neq 0.\label{condition:invert}
\end{align}
Since the disformal transformation contains derivatives of $\phi$,
the theory transformed from Horndeski
has higher-order field equations.\footnote{If both $C$ and $D$ depend only on $\phi$,
the transformed field equations remain of second order and so
the Horndeski theory is mapped to Horndeski~\cite{Bettoni:2013diz}.}
Nevertheless, it is a degenerate theory
with $(2+1)$ degrees of freedom
because
an invertible field redefinition does not change
the number of physical degrees of
freedom~\cite{Arroja:2015wpa,Domenech:2015tca,Takahashi:2017zgr}.
This example implies the existence of a wider class
of healthy scalar-tensor theories than the Horndeski class.


Degenerate higher-order scalar-tensor
theories have been constructed and investigated systematically
in~\cite{Langlois:2015cwa,Langlois:2015skt,Crisostomi:2016czh,%
Achour:2016rkg,BenAchour:2016fzp,Langlois:2017mxy}.
Let us follow Ref.~\cite{Langlois:2015cwa}
and consider the extension of
Horndeski's $G_4$ Lagrangian given by
\begin{align}
{\cal L}=f(\phi,X)R+\sum_{I=1}^5A_I(\phi,X)L_I,\label{qDHOSTL1}
\end{align}
where
\begin{align}
&L_1=\phi_{\mu\nu}\phi^{\mu\nu},\quad L_2=(\Box\phi)^2,
\quad L_3=\Box\phi \phi^\mu\phi^\nu\phi_{\mu\nu}
\notag \\ &
L_4=\phi^\mu\phi_{\mu\alpha}\phi^{\alpha \nu}\phi_\nu,
\quad
L_5=(\phi^\mu\phi^\nu\phi_{\mu\nu})^2.
\end{align}
These five constituents exhaust all the possible quadratic terms in
second derivatives of $\phi$,
and the Horndeski theory is the special case with
$A_2=-A_1=f_X$ and $A_3=A_4=A_5=0$.
The scalar field (respectively, the metric)
corresponds to $\phi$ (respectively, $q$) in the previous
mechanical toy model.
By inspecting the structure of the highest derivative terms
in~\eqref{qDHOSTL1},\footnote{We require the degeneracy
in any coordinate systems. It is argued in~\cite{DeFelice:2018mkq} that one can
relax this requirement and consider theories that are degenerate when
restricted to the unitary gauge.}
one finds that the degeneracy conditions are given by three equations
relating the six functions in the Lagrangian, leaving
three arbitrary functions (except for some special cases).
The degenerate theories whose Lagrangian is of the form~\eqref{qDHOSTL1}
are called quadratic DHOST theories.
Note that one is free to add to~\eqref{qDHOSTL1} the Horndeski terms
$G_2(\phi,X)-G_3(\phi,X)\Box\phi$,
because these two terms are nothing to do with
the degeneracy conditions.

Quadratic DHOST theories are classified into several
subclasses~\cite{Langlois:2015cwa,Crisostomi:2016czh,Achour:2016rkg}.
Of particular importance among them is the so called class Ia,
which is characterized by
\begin{align}
A_2&=-A_1,
\\
 A_4&=\frac{1}{2(f+2XA_1)^2}[
 8XA_1^3+(3f+16Xf_X)A_1^2-X^2fA_3^2+2X(4Xf_X-3f)A_1A_3
 \notag \\ &\qquad\qquad\qquad\quad
 +2f_X(3f+4Xf_X)A_1+2f(Xf_X-f)A_3+3ff_X^2
 ],
 \\
 A_5&=
-\frac{\left(f_X+A_1+XA_3\right)
\left(2fA_3-f_XA_1-A_1^2+3XA_1A_3\right)}{2(f+2XA_1)^2}
 ,
\end{align}
with $f+2XA_1\neq 0$.
(Recall that we are using the notation $X:=-g^{\mu\nu}\phi_\mu\phi_\nu/2$.)
The arbitrary functions are thus taken to be
$f$, $A_1$, and $A_3$.
Cosmology in this class of DHOST theories has been
studied in Refs.~\cite{Crisostomi:2017pjs,Frusciante:2018tvu,Crisostomi:2018bsp},
where it is demonstrated that the apparently higher-order
equations of motion can be reduced to a second-order system
for the scale factor and the scalar field.

Clearly, the Horndeski theory is included in class Ia.
Another important particular case is (a subclass of) the
GLPV theory~\cite{Gleyzes:2014dya,Gleyzes:2014qga}
satisfying
\begin{align}
A_2=-A_1=f_X+XA_3\quad \Rightarrow\quad  A_4=-A_3, \quad A_5=0.
\end{align}
In this case one has two arbitrary functions $f$ and $A_3$,
and the Horndeski theory is reproduced by further taking $A_3=0$.
The Lagrangian for the GLPV theory is written explicitly as
\begin{align}
{\cal L}_{\rm GLPV}&=fR+f_X\left[(\Box\phi)^2-\phi_{\mu\nu}\phi^{\mu\nu}\right]
\notag \\ &\quad
+A_3\left\{
X\left[(\Box\phi)^2-\phi_{\mu\nu}\phi^{\mu\nu}\right]
+\Box\phi \phi^\mu\phi^\nu\phi_{\mu\nu}-
\nabla_\mu X\nabla^\mu X
\right\}.\label{GLPV4L}
\end{align}
Interestingly, the second line (with $A_3=\,$const) can be obtained
from a naive, minimal covariantization of the Galileon theory
(see the second line in Eq.~\eqref{eq:Lag_original_Galileon1}).
In Sec.~\ref{subsec:GtoH} we introduced the counter term
to cancel the higher derivatives which appear upon covariantization.
However, this example shows that
without the counter term we still have a healthy degenerate theory~\cite{Deffayet:2015qwa}.

A notable property of DHOST theories in class Ia is that
all the Lagrangians can be mapped into a Horndeski Lagrangian
through a disformal transformation~\eqref{def:disformalTr}~\cite{Achour:2016rkg}.
In other words, one can remove two of the three functions of $\phi$ and $X$
in the quadratic DHOST sector by using the two functions, $C(\phi,X)$
and $D(\phi,X)$, in the disformal transformation,
to move into a ``Horndeski frame'' with a single function $G_4(\phi,X)$
at quadratic order. At this point
it is worth emphasizing that
class Ia DHOST theories in the presence of minimally coupled matter
are equivalent to Horndeski
with disformally coupled matter, but {\em not} to Horndeski with
minimally coupled matter. This fact is crucial
in particular to the screening mechanism discussed in the next section.

Subclasses other than class Ia are phenomenologically unacceptable.
In these subclasses, the gradient terms in
the quadratic actions for scalar and tensor cosmological perturbations
have opposite signs (i.e., either of the two modes is unstable),
or tensor perturbations are
nondynamical~\cite{Langlois:2017mxy,deRham:2016wji}.
Therefore, only the DHOST theories disformally related to Horndeski
can be viable.

In this subsection we have focused for simplicity on DHOST theories whose
Lagrangian is a quadratic polynomial in $\phi_{\mu\nu}$.
One can do similar manipulation to construct cubic DHOST theories
as an extension of Horndeski's $G_5$ Lagrangian
(i.e., DHOST theories whose Lagrangian is a cubic polynomial in
$\phi_{\mu\nu}$)~\cite{BenAchour:2016fzp},
though their classification is much more involved.
Cubic DHOST theories disformally disconnected to Horndeski
also exhibit gradient instabilities in tensor or scalar modes.

One can go beyond the polynomial assumption and
generate a novel family of DHOST theories from
nondegenerate theories via a noninvertible
disformal transformation with
\begin{align}
D(\phi,X)=\frac{C(\phi,X)}{2X}+F(\phi),
\end{align}
where $C(\phi,X)$ and $F(\phi)$ are arbitrary
(see Eq.~\eqref{condition:invert})~\cite{Takahashi:2017pje,Langlois:2018jdg}.
This is essentially the field redefinition used
in the context of mimetic gravity~\cite{Chamseddine:2013kea,Chamseddine:2014vna}
(see~\cite{Sebastiani:2016ras} for a review).
The idea behind this is that
noninvertible field redefinition can change the number of
dynamical degrees of freedom.
New DHOST theories thus generated are not disformally connected to Horndeski in general.
Such ``mimetic DHOST'' theories suffer from
gradient instabilities of tensor or scalar modes~\cite{Takahashi:2017pje,Langlois:2018jdg}
(see also~\cite{Ramazanov:2016xhp,Ganz:2018mqi} for more about
this instability issue).

The idea of degenerate theories can be extended to
include more than one higher derivative fields~\cite{Motohashi:2016ftl,Crisostomi:2017aim},
though it seems challenging to construct
concrete nontrivial examples of a multi-scalar version of DHOST theories.
Degenerate theories involving only the metric
were explored in~\cite{Crisostomi:2017ugk} under the name of
``beyond Lovelock gravity.''

\subsection{After GW170817}

Measuring the speed of gravitational waves $c_{\rm GW}$
can be a test for modified gravity
theories~\cite{Nishizawa:2014zna,Lombriser:2015sxa,Lombriser:2016yzn,Bettoni:2016mij}.
Indeed, the nearly simultaneous detection of gravitational waves GW170817 and the
$\gamma$-ray burst GRB 170817A~\cite{TheLIGOScientific:2017qsa,Monitor:2017mdv,GBM:2017lvd}
provides a tight constraint on $c_{\rm GW}$ and hence on
scalar-tensor theories and other types of modified gravity.
The limit on the difference between $c_{\rm GW}$ and the speed of light
imposed by this recent event is\footnote{The
lower bound on $c_{\rm GW}$ can also be obtained from the argument on the
gravitational Cherenkov radiation, which can even be tighter than
this~\cite{Moore:2001bv,Kimura:2011qn}.
However, the frequencies concerned are much higher than those of LIGO observations.}
\begin{align}
-3\times 10^{-15} < c_{\rm GW}-1 < 7\times 10^{-16}.
\end{align}
This constraint motivates us to identify the viable subclass of the Horndeski and DHOST
theories as an alternative to dark energy satisfying
$c_{\rm GW}=1$~\cite{Creminelli:2017sry,Sakstein:2017xjx,Ezquiaga:2017ekz,Baker:2017hug,%
Bartolo:2017ibw,Kase:2018iwp,Ezquiaga:2018btd,Kase:2018aps}
(see also~\cite{Amendola:2018ltt,Copeland:2018yuh}
for scalar-tensor theories that achieve $c_{\rm GW}=1$ dynamically).

We start with the Horndeski theory, in which the general form of
the propagation speed of gravitational waves is given by Eq.~\eqref{speedofGW_H}:
\begin{align}
c_{\rm GW}^2 =
 \frac{G_4-X(\ddot\phi G_{5X}+G_{5\phi})}{G_4-2XG_{4X}-X(H\dot\phi G_{5X}-G_{5\phi})}.
\end{align}
In order for this to be equal to the speed of light
irrespective of the background cosmological evolution,
we require that
\begin{align}
G_{4X}=0, \quad G_5=0.
\end{align}
Thus, the viable subclass within Horndeski is described by the Lagrangian
\begin{align}
{\cal L}=G_2(\phi,X)-G_{3}(\phi,X)\Box\phi + G_4(\phi)R.
\end{align}
This excludes for instance
the scalar field coupled to the Gauss-Bonnet term.

Let us then consider DHOST theories. It turns out that
any term in the cubic DHOST Lagrangians leads to $c_{\rm GW}\neq 1$
(just as the $G_5$ term in the Horndeski theory does),
and hence all cubic DHOST theories are ruled out.
In quadratic DHOST theories,
the action for the tensor perturbations is given by~\cite{Langlois:2017mxy,deRham:2016wji}
\begin{align}
  S^{(2)}_{\rm tensor}=
  \frac{1}{4}\int\D t\D^3x a^3\left[
  (f+2XA_1)\dot h_{ij}^2-\frac{f}{a^2}(\partial_k h_{ij})^2
  \right].\label{tensor2_dhost}
\end{align}
From this we see that the propagation speed of gravitational waves is
\begin{align}
c_{\rm GW}^2=\frac{f}{f+2XA_1},
\end{align}
and so we impose $A_1=0$. The viable subclass in DHOST theories
thus reduces to
\begin{align}
A_2&=-A_1=0,\label{c1t1}
\\
A_4&=\frac{1}{2f}[
-X^2A_3^2+2(Xf_X-f)A_3+3f_X^2
],\label{c1t2}
\\
A_5&=
-\frac{A_3\left(f_X+XA_3\right)}{f}.\label{c1t3}
\end{align}
We have two free functions $f(\phi,X)$ and $A_3(\phi,X)$
in addition to the lower order Horndeski terms $G_2(\phi,X)$ and $G_3(\phi,X)$.

More recently, it was pointed out that
gravitons can decay into $\phi$ in DHOST theories~\cite{Creminelli:2018xsv}.
To avoid this graviton decay, it is further required that $A_3=0$.
(Otherwise, gravitational waves would not be observed.)
We thus finally have
\begin{align}
A_4=\frac{3f_X^2}{2f},\quad A_1=A_2=A_3=A_5=0.
\end{align}
Note that this subclass does not belong to Horndeski nor GLPV families (if $f_X\neq 0$).

As argued in the previous subsection, mimetic gravity can be viewed as
a kind of DHOST theories.
The implications of GW170817 for the mimetic class of DHOST theories
have been discussed in~\cite{Casalino:2018tcd,Ganz:2018vzg,Casalino:2018wnc}.

It should be emphasized that in constraining scalar-tensor theories
with gravitational waves
we have assumed that the DHOST theory under consideration
 as an alternative description of
dark energy is valid on much higher energy scales where
LIGO observations are made ($\sim 100\,$Hz $\sim 10^{-13}\,$eV).
The validity of this assumption needs to be looked
into carefully~\cite{deRham:2018red}.
Similarly, gravity at much higher energies than this
remains unconstrained. Therefore,
modified gravity in the early universe is free from
these gravitational wave constraints.

A final remark is that, even if $c_{\rm GW}=1$, $f$ (or $G_4$)
may depend on time, which gives rise to the extra contribution $\dot f/f$ to
the friction term in the equation of motion for $h_{ij}$.
This results in a modification of the amplitude of gravitational
waves, which can be measurable~\cite{Belgacem:2017ihm,Amendola:2017ovw,Nunes:2018zot}.

\section{Vainshtein screening}\label{sec:Vainshtein_mech}

While a scalar-tensor theory as an alternative to dark energy
is supposed to give rise to
${\cal O}(1)$ modification of gravity on cosmological scales,
the extra force mediated by the scalar degree of freedom $\phi$
must be screened on small scales where
general relativity has been tested to high precision.
This occurs if $\phi$ is effectively massive
in the vicinity of a source, or if $\phi$ is effectively
weakly coupled to the source.
The former case corresponds to the chameleon
mechanism~\cite{Khoury:2003aq,Khoury:2003rn},
in which the effective potential for $\phi$ depends on
the local energy density through the coupling of $\phi$ to matter.
The latter case is based on the idea of Vainshtein~\cite{Vainshtein:1972sx}
(see also~\cite{Deffayet:2001uk}) and is called the Vainshtein mechanism.
(There are other screening mechanisms called
symmetron~\cite{Hinterbichler:2010es,Hinterbichler:2011ca}
and k-Mouflage models~\cite{Babichev:2009ee}, both of which effectively
suppress the coupling to matter.)
The Vainshtein mechanism is relevant to the Galileon theories, and
below we will review this screening mechanism
in the context of the generalized Galileon/Horndeski theory.
See also~\cite{Babichev:2013usa} for a nice review on the
Vainshtein mechanism.


\subsection{A Vainshtein primer}

We start with emphasizing the need for a screening mechanism,
and then introduce the Vainshtein mechanism.

To see how gravity is modified around matter
in a simple model and as a result
it fails to satisfy the experimental constraints,
let us consider a theory
\begin{align}
S=\int\D^4x\sqrt{-g}\left[f(\phi)R
+X\right]
 +S_m[g_{\mu\nu}, \psi_m],\label{action:vainshtein0}
\end{align}
where the matter fields (denoted as $\psi_{m}$) are minimally
coupled to the metric $g_{\mu\nu}$.

We investigate perturbations around
a Minkowski background with a constant scalar $\phi$,
\begin{align}
g_{\mu\nu}=\eta_{\mu\nu}+\mpl^{-1}h_{\mu\nu}(t,\Vec{x}),
\quad
\phi =\phi_0+\varphi(t,\Vec{x}),\label{def:pert-vainshtein}
\end{align}
caused by the energy-momentum tensor for matter, $T_{\mu\nu}$
(the above theory admits the background solution
$g_{\mu\nu}=\eta_{\mu\nu}$ and $\phi=\phi_0=\,$const).
Here we write $f(\phi_0)=\mpl^2/2$ and
defined the metric perturbations so that
$h_{\mu\nu}$ has the dimension of mass.
Expanding~\eqref{action:vainshtein0} to second order in perturbations,
we obtain the effective Lagrangian for the description of
weak gravitational fields as
\begin{align}
{\cal L}_{\rm eff}=
-\frac{1}{4}h^{\mu\nu}\hat {\cal E}_{\mu\nu}^{\alpha\beta}h_{\alpha\beta }
-\frac{1}{2}\partial_\mu\varphi\partial^\mu\varphi -\xi h^{\mu\nu}X_{\mu\nu}^{(1)}+
\frac{1}{2\mpl }h^{\mu\nu}T_{\mu\nu},\label{action:vainshtein01}
\end{align}
where $\xi:=\mpl^{-1}\D f/\D\phi|_{\phi=\phi_0}$,
\begin{align}
X_{\mu\nu}^{(1)}&:=\eta_{\mu\nu}\Box\varphi-\varphi_{\mu\nu},\label{def:X1mn}
\end{align}
and
\begin{align}
\hat {\cal E}_{\mu\nu}^{\alpha\beta}h_{\alpha\beta }:=-\frac{1}{2}\Box h_{\mu\nu}
+\partial^\lambda\partial_{(\mu}h_{\nu)\lambda}+\frac{1}{2}\eta_{\mu\nu}\Box h
-\frac{1}{2}\eta_{\mu\nu}\partial_\lambda\partial_\rho h^{\lambda\rho}
-\frac{1}{2}\partial_\mu\partial_\nu h
\end{align}
is the linearized Einstein tensor (divided by $\mpl$).
Here, indices are raised and lowered by $\eta_{\mu\nu}$.

The third term in the Lagrangian~\eqref{action:vainshtein01}
signals the mixing of the scalar degree of freedom with the graviton.
This can be disentangled by making use of the field redefinition
\begin{align}
h_{\mu\nu}=\tilde h_{\mu\nu} -2\xi\varphi\eta_{\mu\nu},
\label{tr:conformal1}
\end{align}
leading to
\begin{align}
{\cal L}_{\rm eff}=
-\frac{1}{4}\tilde h^{\mu\nu}\hat {\cal E}_{\mu\nu}^{\alpha\beta}\tilde h_{\alpha\beta }
-\frac{1+6\xi^2}{2}\partial_\mu\varphi\partial^\mu\varphi +
\frac{1}{2\mpl }\tilde h^{\mu\nu}T_{\mu\nu}
-\frac{\xi}{\mpl}\varphi T
.\label{action:vainshtein02}
\end{align}
The transformation~\eqref{tr:conformal1} is equivalent to
the linear part of
the conformal transformation to the Einstein frame,
$\tilde g_{\mu\nu}=C(\phi)g_{\mu\nu}$
with $C= f(\phi)/f(\phi_0)$.
In the new frame we have the nonminimal coupling of the form $\varphi T$,
and the field equations are given by
\begin{align}
\hat {\cal E}_{\mu\nu}^{\alpha\beta}\tilde h_{\alpha\beta }&=\mpl^{-1}T_{\mu\nu},
\\
(1+6\xi^2)\Box\varphi &= \mpl^{-1}\xi T.
\end{align}
Thus, ${\cal O}(1)$ modification of gravity is expected for $\xi={\cal O}(1)$.

To be more concrete, let us consider a spherical distribution of
nonrelativistic matter, $T_{\mu\nu}=\rho(r)\delta_\mu^0\delta_\nu^0$,
with $\tilde h_{00}=-2\tilde \Phi(r)$ and $\tilde h_{ij}=-2\tilde\Psi(r)\delta_{ij}$.
Then, the field equations read
\begin{align}
\frac{1}{r^2}\left(r^2\tilde\Psi'\right)'&=\frac{\rho}{2\mpl},
\\
\tilde\Psi-\tilde\Phi&=0,
\\
\frac{1}{r^2}\left(r^2\varphi'\right)'&=-\frac{\xi}{1+6\xi^2}\frac{\rho}{\mpl},
\end{align}
where a prime stands for differentiation with respect to $r$.
These equations can be integrated straightforwardly to give
\begin{align}
\mpl^{-1}\tilde\Phi'=\mpl^{-1}\tilde\Psi' = (8\pi\mpl^2)^{-1}\frac{{\cal M}(r)}{r^2},
\quad
\mpl^{-1}\varphi'=-
\frac{2\xi}{1+6\xi^2}\cdot
(8\pi\mpl^2)^{-1}\frac{{\cal M}(r)}{r^2},
\label{sol:linear1}
\end{align}
where ${\cal M}(r)$ is the enclosed mass, ${\cal M}(r):=4\pi \int^r\rho(s)s^2\D s$.
It follows from~\eqref{tr:conformal1} that the metric perturbations
in the original frame are given by
$\Phi = \tilde\Phi-\xi\varphi$ and $\Psi=\tilde\Psi+\xi\varphi$.
Thus, the metric potentials outside the matter distribution are given by
\begin{align}
  \Phi = -\frac{G_N{\cal M}}{r},\quad
  \Psi =\gamma \Phi,
\end{align}
with
\begin{align}
8\pi G_N:=\frac{1+8\xi^2}{1+6\xi^2}\frac{1}{\mpl^2},
\quad
\gamma-1=-\frac{4\xi^2}{1+8\xi^2}.
\end{align}
For $\xi={\cal O}(1)$ we have $\gamma-1={\cal O}(1)$, which clearly
contradicts the solar-system experiments~\cite{Will:2014kxa}.

Now we add a Galileon-like cubic interaction to~\eqref{action:vainshtein02}:
\begin{align}
  {\cal L}_{\rm eff}=
  -\frac{1}{4}\tilde h^{\mu\nu}\hat {\cal E}_{\mu\nu}^{\alpha\beta}\tilde h_{\alpha\beta }
  -\frac{1+6\xi^2}{2}(\partial\varphi)^2
-\frac{1}{2\Lambda^3}(\partial\varphi)^2\Box\varphi
   +
  \frac{1}{2\mpl }\tilde h^{\mu\nu}T_{\mu\nu}
  -\frac{\xi}{\mpl}\varphi T.
  \label{action:cubic-vainshtein}
\end{align}
Then, the scalar-field equation of motion becomes
\begin{align}
(1+6\xi^2)\Box\varphi +\frac{1}{\Lambda^3}
\left[(\Box\varphi)^2-\varphi_{\mu\nu}\varphi^{\mu\nu}\right]=\frac{\xi}{\mpl}T,
\end{align}
which, for a spherical matter distribution such as a star, gives
\begin{align}
(1+6\xi^2)r^2\varphi' + \frac{2}{\Lambda^3}r(\varphi')^2
 = -\frac{\xi {\cal M}(r)}{4\pi \mpl}.
\end{align}
This equation can be solved algebraically, yielding
\begin{align}
\varphi'=
c \Lambda^3  r
\left[
-1+\sqrt{1-\frac{\xi}{c^2}\left(\frac{r_V}{r}\right)^3}
\right],
\label{sol:vainshtein1}
\end{align}
where $c:=(1+6\xi^2)/4$ is an ${\cal O}(1)$ constant
and we defined
\begin{align}
r_V:=\left(\frac{{\cal M}}{8\pi \mpl\Lambda^3}\right)^{1/3}.\label{def:vrad}
\end{align}
(We consider a stellar exterior so that now ${\cal M}(=\,$const) is the
mass of the star.)
For $r\gg r_V$, Eq.~\eqref{sol:vainshtein1} reproduces~\eqref{sol:linear1}.
However, for $r\ll r_V$, we find
\begin{align}
\varphi'\simeq (-\xi)^{1/2}\left(\frac{r}{r_V}\right)^{3/2}\tilde\Phi'\ll \tilde\Phi'
\quad\Rightarrow\quad
\frac{\Phi}{\mpl}\simeq \frac{\Psi}{\mpl}
\simeq -\frac{G_N{\cal M}}{r},\quad 8\pi G_N:=\frac{1}{\mpl^2}.
\end{align}
It turns out that
the nonlinear interaction introduced in~\eqref{action:cubic-vainshtein}
helps the recovery of standard gravity, and
the solar-system constraints can thus be evaded
if $r_V$ is sufficiently large.
This is the {\em Vainshtein mechanism}, and
$r_V$ is called the {\em Vainshtein radius},
within which general relativity is reproduced.
Although we are considering small perturbations,
we see that
\begin{align}
\frac{\Box\varphi}{\Lambda^3}\gtrsim{\cal O}(1)\quad {\rm for}\quad r\lesssim r_V.
\label{O1_second_deri}
\end{align}
This tells us why nonlinearity is important even in a weak gravity environment.

If the scalar degree of freedom accounts for the present
accelerating expansion of the universe, $\Lambda$ is expected to be
as small as
\begin{align}
\Lambda \sim (\mpl H_0^2)^{1/3},\label{estimate:Lambda}
\end{align}
where $H_0$ is the present Hubble scale.
This is deduced from the estimate
\begin{align}
\mpl^2 H^2_0\sim \dot\phi^2\sim \frac{\dot\phi^2\ddot\phi}{\Lambda^3},
\quad \ddot\phi\sim H_0\dot\phi.
\end{align}
For $M\sim M_{\odot}$, Eq.~\eqref{def:vrad} with~\eqref{estimate:Lambda} gives
\begin{align}
r_V\sim 100\,{\rm pc},
\end{align}
which is much larger than the size of the solar system.

\subsection{Vainshtein screening in Horndeski theory}

We can repeat the same analysis in the Horndeski
theory~\cite{Narikawa:2013pjr,Koyama:2013paa,DeFelice:2011th,Kase:2013uja}.
We only consider the case with $G_5=0$ for a reason to be
explained later.
In order for the background $g_{\mu\nu}=\eta_{\mu\nu}$
with $\phi=\phi_0=\,$const is a solution, we require that
$G_2(\phi_0,0)=G_{2\phi}(\phi_0,0)=0$.

In substituting~\eqref{def:pert-vainshtein}
to the Horndeski action
(now $\mpl$ is defined by $G_{4}(\phi_0,0)=\mpl^2/2$)
and expanding it in terms of
perturbations, one must carefully retain
the nonlinear terms with second derivatives
because they can be large on small scales
as suggested by~\eqref{O1_second_deri}.
More specifically, we have the terms of the following forms in the Lagrangian:
\begin{align}
(\partial h_{\mu\nu})^2,\quad (\partial\varphi)^2,
\quad (\partial\varphi)^2(\partial^2\varphi)^n,\quad h_{\mu\nu}(\partial^2\varphi)^n.
\end{align}
However, as we are interested in the Vainshtein mechanism,
we ignore the mass term $K_{\phi\phi}\varphi^2$.
We thus find (in the original frame)~\cite{Koyama:2013paa}
\begin{align}
{\cal L}_{\rm eff}&=
-\frac{1}{4}h^{\mu\nu}\hat {\cal E}_{\mu\nu}^{\alpha\beta} h_{\alpha\beta }
-\frac{\eta}{2}(\partial\varphi)^2
+\frac{\mu}{\Lambda^3}{\cal L}_3^{\rm Gal}+\frac{\nu}{\Lambda^6}{\cal L}_4^{\rm Gal}
\notag \\ &\quad
-\xi h^{\mu\nu}X_{\mu\nu}^{(1)}-\frac{\alpha}{\Lambda^3}h^{\mu\nu}X_{\mu\nu}^{(2)}
+\frac{1}{2\mpl}h^{\mu\nu}T_{\mu\nu},
\label{effLag:Vainshtein1}
\end{align}
where
\begin{align}
{\cal L}_3^{\rm Gal}&:=-\frac{1}{2}(\partial\varphi)^2\Box\varphi,
\\
{\cal L}_4^{\rm Gal}&:=-\frac{1}{2}(\partial\varphi)^2\left[
(\Box\varphi)^2-\varphi_{\mu\nu}\varphi^{\mu\nu}
\right],
\end{align}
$X_{\mu\nu}^{(1)}$ was already defined in Eq.~\eqref{def:X1mn}, and
\begin{align}
X_{\mu\nu}^{(2)}:=\varphi_\mu^\alpha \varphi_{\alpha\nu}-\Box\varphi \varphi_{\mu\nu}
+\frac{1}{2}\eta_{\mu\nu}\left[(\Box\varphi)^2-\varphi_{\alpha\beta}\varphi^{\alpha\beta}
\right ].
\end{align}
We have defined
the dimensionless parameters $\eta$,
$\xi$, $\mu$, $\nu$, and $\alpha$ by
\begin{align}
&G_{4\phi}=\mpl \xi ,\quad G_{2X}-2G_{3\phi}=\eta,
\quad
G_{3X}-3G_{4\phi X}=-\frac{\mu}{\Lambda^3}
\notag \\&
G_{4X}=\frac{\mpl\alpha}{\Lambda^3},
\quad
G_{4XX}=\frac{\nu}{\Lambda^6},
\end{align}
with $\Lambda$ being some energy scale.
These dimensionless parameters are assumed to be ${\cal O}(1)$
unless they vanish.
The Lagrangian~\eqref{effLag:Vainshtein1}
describes the effective theory for the Vainshtein mechanism.
Note that this effective theory
has the Galilean shift symmetry, $\varphi\to \varphi + b_\mu x^\mu + c$.

One notices the presence of the
new term representing the mixing of the scalar degree of freedom
and the graviton: $h^{\mu\nu}X_{\mu\nu}^{(2)}$.
This, as well as $h^{\mu\nu}X_{\mu\nu}^{(1)}$, can be demixed
through the field redefinition~\cite{deRham:2010tw}
\begin{align}
h_{\mu\nu}=\tilde h_{\mu\nu}-2\xi\varphi \eta_{\mu\nu} +\frac{2\alpha}{\Lambda^3}
\partial_\mu\varphi\partial_\nu\varphi.
\end{align}
The new piece $(2\alpha/\Lambda^3)\partial_\mu\varphi\partial_\nu\varphi$
is equivalent to a disformal transformation.
After this transformation
the effective Lagrangian~\eqref{effLag:Vainshtein1}
reduces to
\begin{align}
  {\cal L}_{\rm eff}&=
  -\frac{1}{4}\tilde h^{\mu\nu}\hat {\cal E}_{\mu\nu}^{\alpha\beta}
  \tilde h_{\alpha\beta }
  -\frac{\eta+6\xi^2}{2}(\partial\varphi)^2
  +\frac{\mu+6\alpha\xi}{\Lambda^3}{\cal L}_3^{\rm Gal}
  +\frac{\nu+2\alpha^2}{\Lambda^6}{\cal L}_4^{\rm Gal}
  \notag \\ &\quad
  +\frac{1}{2\mpl}\tilde h^{\mu\nu}T_{\mu\nu}-\frac{\xi}{\mpl}\varphi T
  +\frac{\alpha}{\mpl\Lambda^3}\partial_\mu\varphi\partial_\nu\varphi T^{\mu\nu}.
  \label{effLag:Vainshtein2}
\end{align}
Things are more transparent in this Einstein frame than in the
original Jordan frame.

To see how the Vainshtein mechanism operates generically,
let us again consider a spherically symmetric matter distribution.
The field equation for $\varphi$,
\begin{align}
&(\eta+6\xi^2)\Box\varphi
+\frac{\mu+6\alpha\xi}{\Lambda^3}\left[
(\Box\varphi)^2-\varphi_{\mu\nu}\varphi^{\mu\nu}
\right]
\notag \\ &
+\frac{\nu+2\alpha^2}{\Lambda^6}
\left[
(\Box\varphi)^3-3\varphi_{\mu\nu}\varphi^{\mu\nu}\Box\varphi
+2 \varphi_{\mu\nu}\varphi^{\nu\lambda}\varphi_\lambda^\mu
\right]
=\frac{\xi}{\mpl}T+\frac{2\alpha}{\mpl}\varphi_{\mu\nu}T^{\mu\nu},
\end{align}
can be written in the following form after integrated once:
\begin{align}
\frac{\eta+6\xi^2}{2}x+(\mu+6\alpha\xi)x^2+(\nu+2\alpha^2)x^3=-\xi A,
\label{poly:vain}
\end{align}
where we introduced the convenient dimensionless quantities
\begin{align}
x(r):=\frac{1}{\Lambda^3}\frac{\varphi'}{r},
\quad
A(r):=\frac{1}{\mpl\Lambda^3}\frac{{\cal M}(r)}{8\pi r^3}.
\end{align}
The field equations for $\tilde h_{\mu\nu}$ imply
\begin{align}
  \frac{1}{\Lambda^3}\frac{\tilde \Phi'}{r}
=\frac{1}{\Lambda^3}\frac{\tilde \Psi'}{r} = A.
\end{align}
Since the special case with $\nu+2\alpha^2=0$ was already essentially analyzed
in the previous subsection, we focus on the generic case
with $\nu+2\alpha^2\neq 0$.
We have, for $A\gg 1$,
\begin{align}
x\simeq \left(\frac{-\xi A}{\nu+2\alpha^2}\right)^{1/3}.\label{soln:x_vainshtein}
\end{align}
The metric perturbations in the Jordan frame are obtained from
$\Phi=\tilde\Phi-\xi\varphi$ and $\Psi=\tilde\Psi + \xi\varphi -(\alpha/\Lambda^3)(\varphi')^2$,
but we see from~\eqref{soln:x_vainshtein} that
the extra scalar-field contributions are small, yielding
$\mpl^{-1}\Phi'\simeq\mpl^{-1}\Psi' \simeq G_N{\cal M}/r^2$
where $8\pi G_N=\mpl^{-2}=[2 G_4(\phi_0,0)]^{-1}$.
Since $A\propto r^{-3}$ outside the source,
it is appropriate to define the Vainshtein radius
$r_V:=({\cal M}/8\pi \mpl\Lambda^3)^{1/3}$
so that $A=(r_V/r)^3$.
The nonlinearity in the scalar-field equation of motion thus
helps to suppress the force mediated by $\varphi$, so that
standard gravity is recovered inside the Vainshtein radius $r_V$.

Though the expression is slightly more complicated,
the complete effective Lagrangian from
the Horndeski theory including the $G_5$ term can be obtained
in the same way as above~\cite{Koyama:2013paa}.
One then finds the quintic Galileon interaction for $\varphi$
and another mixing term between $\varphi$ and $h_{\mu\nu}$
in the effective Lagrangian.
This mixing cannot be eliminated by a field redefinition~\cite{deRham:2010tw}.
It can be shown that the screened region outside the spherically symmetric
matter distribution is unstable against linear perturbations
in the presence of this mixing~\cite{Koyama:2013paa}.

So far we have considered the simplest background solution,
$g_{\mu\nu}=\eta_{\mu\nu}$ with $\phi=\phi_0=\,$const,
and static, spherically symmetric perturbations on top of the background.
The above analysis has been extended to a cosmological background
with time-dependent $\phi_0$ in~\cite{Kimura:2011dc}.
On a cosmological background
we start with the Newtonian gauge metric~\eqref{metric:Newtonian}
rather than~\eqref{def:pert-vainshtein}
because Lorentz invariance is spontaneously broken due to
nonvanishing $\dot\phi_0(t)$.
Keeping the appropriate nonlinear terms,
the Lagrangian under the quasi-static approximation is
computed as
\begin{align}
{\cal L}_{\rm eff}&=
a\left[M^2\left(-c_{\rm GW}^2\Psi\partial^2\Psi +2\Psi\partial^2\Phi\right)
-\frac{\eta}{2}(\partial\varphi)^2 -2M\left(
\xi_1\Phi -2\xi_2\Psi\right)X^{(1)}
\right]
\notag \\ & \quad +
\frac{\mu}{a\Lambda^3}{\cal L}_{3}^{\rm Gal}+
\frac{\nu}{a^3\Lambda^6}{\cal L}_{4}^{\rm Gal}
-\frac{2M}{a\Lambda^3}\left(\alpha_1\Phi-\alpha_2\Psi\right)X^{(2)}-a^3 \Phi\delta \rho,
\label{effL:vain_cos}
\end{align}
where $\delta\rho$ is a density perturbation,
\begin{align}
M^2:={\cal G}_T,
\quad
X^{(1)}:=\partial^2\varphi,
\quad
X^{(2)}:=\frac{1}{2}\left[(\partial^2\varphi)^2-\partial_i\partial_j\varphi
\partial^i\partial^j\varphi\right],
\end{align}
and
\begin{align}
{\cal L}_3^{\rm Gal}:=-\frac{1}{2}(\partial\varphi)^2 \partial^2\varphi,
\quad
{\cal L}_4^{\rm Gal}:=-\frac{1}{2}(\partial\varphi)^2\left[
( \partial^2\varphi)^2 -\partial_i\partial_j\varphi
\partial^i\partial^j\varphi\right].
\end{align}
The coefficients are (slowly-varying) functions of time in general.
Explicitly, we have
\begin{align}
\frac{M\alpha_1}{\Lambda^3}:=G_{4X}+2XG_{4XX},
\quad
\frac{M\alpha_2}{\Lambda^3}:=G_{4X},
\quad
\frac{\nu}{\Lambda^6}:=G_{4XX},
\end{align}
while
\begin{align}
&
M\xi_1\simeq -XG_{3X}+G_{4\phi}+2XG_{4\phi X},
\quad
M\xi_2\simeq G_{4\phi}-2XG_{4\phi X},
\notag \\ &
\frac{\mu}{\Lambda^3}\simeq -\left( G_{3X}-3G_{4\phi X}+2X G_{4\phi XX}\right),
\end{align}
where to simplify the expressions we ignored
$\ddot \phi_0$ and $H$ in $\xi_1$, $\xi_2$, and $\mu$.
The explicit expression for $\eta$ is not important here.
Note that if $c_{\rm GW}^2=1$ then $\alpha_1=\alpha_2=\nu=0$.

In the present case, we stay in the Jordan frame rather than
try to disentangle the couplings between $\varphi$ and
the metric potentials such as $\Phi X^{(1)}$.
For a spherical overdensity, the field equations derived from~\eqref{effL:vain_cos}
are written as
\begin{align}
y-c_{\rm GW}^2z&=-2\xi_2x-\alpha_2 x^2,\label{eq:v_cos1}
\\
z&=A + \xi_1 x+\alpha_1 x^2,\label{eq:v_cos2}
\end{align}
and
\begin{align}
\frac{\eta}{2}x-\xi_1 y + 2\xi_2 z -2 (\alpha_1y-\alpha_2 z)x+\mu x^2+ \nu x^3=0,
\label{eq:v_cos3}
\end{align}
where we introduced
\begin{align}
x:= \frac{1}{\Lambda^3}\frac{\varphi'}{r},
\quad
y:=\frac{M}{\Lambda^3}\frac{\Phi'}{r},
\quad
z:=\frac{M}{\Lambda^3}\frac{\Psi'}{r},
\quad
A:=\frac{1}{M\Lambda^3}\frac{{\cal M}}{8\pi r^3},
\quad
{\cal M}:=4\pi \int^r_0\delta\rho (t,s)s^2\D s,
\end{align}
and took $a\to 1$ for simplicity.
Using Eqs.~\eqref{eq:v_cos1} and~\eqref{eq:v_cos2} one can
remove $y$ and $z$ from Eq.~\eqref{eq:v_cos3} to get
\begin{align}
&\left[c_1+2(\alpha_2-c_{\rm GW}^2\alpha_1)A\right]x
+c_2x^2+\left[
\nu+4\alpha_1\alpha_2-2c_{\rm GW}^2\alpha_1^2
\right]x^3
\notag \\ &
=-(2\xi_2-c_{\rm GW}^2\xi_1)A,\label{poly:vain_cos}
\end{align}
where the coefficients $c_1$ and $c_2$ are written
in terms of $\eta$, $\xi_1$, etc.
This extends Eq.~\eqref{poly:vain} to
a time-dependent background with $\dot\phi_0\neq 0$.

In theories with $c_{\rm GW}^2=1$, Eq.~\eqref{poly:vain_cos}
becomes
\begin{align}
c_1x+c_2x^2=\left(\frac{\eta}{2}-\xi_1^2+4\xi_1\xi_2\right)x+\mu x^2=-(2\xi_2-\xi_1)A.
\end{align}
The Vainshtein radius is defined by $A(r_V)=1$,
and for $A\gg 1$ we have $x\sim A^{1/2}\ll A\simeq y\simeq z$.
Thus, inside the Vainshtein radius the metric potentials
obey
\begin{align}
\Phi'=\Psi'=\frac{G_N{\cal M}}{r^2},\quad
8\pi G_N:=\frac{1}{2G_4(\phi_0(t))}.
\label{timeGN1}
\end{align}
The situation is similar to that for a static background, $\dot\phi_0=0$.
However, it is interesting to see that, even in the minimally-coupled
theories with $G_{4}=\,$const, we still have $M \xi_1\simeq -XG_{3X}\neq 0$
if $\dot\phi_0$ is nonvanishing, so that
$\varphi$ is coupled to the source via the $G_3$ term.


In theories with $c_{\rm GW}^2\neq 1$,
$A$ in the coefficient of the linear term plays an important role.
For $A\gg 1$, Eq.~\eqref{poly:vain_cos} reduces to
\begin{align}
  &2(\alpha_2-c_{\rm GW}^2\alpha_1)Ax
  +\left[
  \nu+4\alpha_1\alpha_2-2c_{\rm GW}^2\alpha_1^2
  \right]x^3\simeq 0
   \notag \\&\Rightarrow \quad
   x^2\simeq
   -\frac{2(\alpha_2-c_{\rm GW}^2\alpha_1)}{\nu+4\alpha_1\alpha_2-2c_{\rm GW}^2\alpha_1^2}A.
   \label{soln:x_v_cos}
\end{align}
This is in contrast to the screened solution
on a static background~\eqref{soln:x_vainshtein}, $x^3\sim A$.
Substituting the solution~\eqref{soln:x_v_cos} to~\eqref{eq:v_cos1}
and~\eqref{eq:v_cos2}, one finds that
\begin{align}
\Phi'=\Psi'=\frac{G_N{\cal M}}{r^2},
\quad
8\pi G_N=\left.\frac{1}{2[G_4-4X(G_{4X}+XG_{4XX})]}\right|_{\phi=\phi_0(t)}.
\label{timeGN2}
\end{align}

As seen from Eqs.~\eqref{timeGN1} and~\eqref{timeGN2},
apparently the standard gravitational law is recovered.\footnote{The
Friedmann equation in the early time takes the form
$3H^2\simeq 8\pi G_{\rm cos}\rho$, where ``cosmological $G$''
coincides with $G$ in Newton's law: $G_{\rm cos}=G_N$.
Note that in general this $G_N$ is different from the effective
gravitational coupling for gravitational waves, $G_{\rm GW}:=(8\pi {\cal F}_T)^{-1}$.
The difference can be constrained from the Hulse-Taylor pulsar~\cite{Jimenez:2015bwa}.}
However,
the effective Newton ``constant'' on a cosmological background
depends on time
even inside the Vainshtein radius
through the cosmological evolution of $\phi_0$~\cite{Babichev:2011iz}.
Although it would be natural to think of
a slow variation $|\dot G_N/G_N|={\cal O}(1)\times H_0 $,
the observational bounds from Lunar Laser Ranging
require a much slower variation,
$|\dot G_N/G_N|< 0.02 H_0 $~\cite{Williams:2004qba}.
This limit can be used to constrain cosmological scalar-tensor theories.

We have seen how Vainshtein screening operates
around a (quasi-)static, spherically symmetric body.
The Vainshtein mechanism away from this simplified setup
has been investigated in~\cite{Brax:2011sv,Hiramatsu:2012xj,deRham:2012fw,%
deRham:2012fg,Chagoya:2014fza,Bloomfield:2014zfa,%
Ogawa:2018srw,Dar:2018dra,Falck:2014jwa,Falck:2015rsa,Falck:2017rvl}.

\subsection{Partial breaking of Vainshtein screening}

In DHOST theories, the operation of the Vainshtein mechanism
turns out to be very nontrivial.
This can be seen as follows.
Let us consider a simple DHOST theory whose Lagrangian
is given by
\begin{align}
{\cal L}&=G_2-G_3\Box\phi +G_4R+G_{4X}\left[(\Box\phi)^2
-\phi_{\mu\nu}\phi^{\mu\nu}\right]
\notag \\ &\quad
+A_3\left\{
X\left[(\Box\phi)^2-\phi_{\mu\nu}\phi^{\mu\nu}\right]
+\Box\phi \phi^\mu\phi^\nu\phi_{\mu\nu}-
\nabla_\mu X\nabla^\mu X
\right\}.
\end{align}
This theory belongs to the GLPV family (see Eq.~\eqref{GLPV4L}).
Due to the new term in the second line,
the effective Lagrangian for the Vainshtein mechanism
under the quasi-static approximation
is now given by~\cite{Kobayashi:2014ida}
(see also~\cite{DeFelice:2015sya,Kase:2015gxi})
\begin{align}
{\cal L}_{\rm eff}&=
a\left\{M^2\left[-c_{\rm GW}^2\Psi\partial^2\Psi +2(1+\alpha_H)\Psi\partial^2\Phi\right]
-\frac{\eta}{2}(\partial\varphi)^2 -2M\left(
\xi_1\Phi -2\xi_2\Psi\right)X^{(1)}
\right\}
\notag \\ & \quad +
\frac{\mu}{a\Lambda^3}{\cal L}_{3}^{\rm Gal}+
\frac{\nu}{a^3\Lambda^6}{\cal L}_{4}^{\rm Gal}
-\frac{2M}{a\Lambda^3}\left(\alpha_1\Phi-\alpha_2\Psi\right)X^{(2)}-a^3 \Phi\delta \rho
\notag \\ &\quad
+\frac{2aM^{3/2}}{\Lambda^{3/2}v}
\alpha_H \dot\Psi
\partial^2\varphi
-\frac{2M}{a\Lambda^3v^2}
\alpha_H
\partial_i\Psi\partial_j\varphi \partial_i\partial_j\varphi,
\label{effL:vain_GLPV}
\end{align}
where
\begin{align}
M^2&:=2\left(G_4-2XG_{4X}-2X^2A_3\right),
\\
\frac{M\alpha_1}{\Lambda^3}&:=G_{4X}+2XG_{4XX}+X(5A_3+2XA_{3X}),
\\
\frac{M\alpha_2}{\Lambda^3}&:=G_{4X}+XA_3,
\\
\frac{\nu}{\Lambda^6}&:=G_{4XX}+2A_3+XA_{3X},
\\
M^2\alpha_H&:=4X^2A_3,
\end{align}
and we write $v:=\dot\phi_0/(M^{1/2}\Lambda^{3/2})\,(={\cal O}(1))$.
Explicit expressions of the other coefficients are not important.
The two terms in the third line are essentially new contributions in
this DHOST theory.
Note that even in the quasi-static regime one cannot, in general,
neglect the first term in the third line because
\begin{align}
\frac{M^{3/2}}{\Lambda^{3/2}v}\alpha_H\dot\Psi\partial^2\varphi
\sim \frac{M^{3/2}H_0}{\Lambda^{3/2}v}\alpha_H\Psi \partial^2\varphi
\sim \frac{M \alpha_H}{v}\Psi X^{(1)}.
\end{align}
This term modifies the linear evolution equation for
density perturbations. More specifically, the coefficient of
the friction term ($\propto \dot\delta$)
in the evolution equation for $\delta$ acquires an additional
contribution other than the Hubble parameter~\cite{Gleyzes:2014dya,Gleyzes:2014qga}.

In the regime where the nonlinear terms are dominant,
one obtains
\begin{align}
(1+\alpha_H)y-c_{\rm GW}^2z&\simeq -\alpha_2 x^2-\frac{\alpha_H}{v^2}
(x^2+rxx'),\label{eq:v_cos_dhost1}
\\
(1+\alpha_H)z&\simeq A +\alpha_1 x^2,\label{eq:v_cos_dhost2}
\end{align}
and
\begin{align}
 -2 (\alpha_1y-\alpha_2 z)x+ \nu x^3-\frac{\alpha_H}{v^2}
\left(3xz+rxz'\right)
 \simeq 0,
\label{eq:v_cos_dhost3}
\end{align}
where for simplicity we ignored the cosmic expansion
by taking $a= 1$.
Equations~\eqref{eq:v_cos_dhost1}--\eqref{eq:v_cos_dhost3}
can be regarded as generalizations of Eqs.~\eqref{eq:v_cos1}--\eqref{eq:v_cos3}.
Using Eqs.~\eqref{eq:v_cos_dhost1} and~\eqref{eq:v_cos_dhost2}
one can express $y$ and $z$ in terms of $x$ and $x'$, and
then eliminate $y$ and $z$ from~\eqref{eq:v_cos_dhost3}.
After doing so
one would obtain a differential equation for $x$. However,
in fact all the derivative terms are canceled out,
yielding an algebraic equation for $x$.
This is the consequence of the degeneracy of the system.
The resultant algebraic equation is solved to give
$x^2 = (\cdots) A+(\cdots) A'$.
Substituting this to Eqs.~\eqref{eq:v_cos_dhost1} and~\eqref{eq:v_cos_dhost2},
we arrive at
\begin{align}
y&=8\pi G_NM^2\left[
A+\frac{\Upsilon_1}{4}\frac{(r^3A)''}{r}
\right],\label{solydhost}
\\
z&=8\pi G_NM^2\left[
A-\frac{5\Upsilon_2}{4}\frac{(r^3A)'}{r^2}
\right],\label{solzdhost}
\end{align}
where we defined the effective Newton ``constant''\footnote{This also
coincides with ``$G_{\rm cos}$'' in the Friedmann equation.}
\begin{align}
8\pi G_N&:=\left[2G_4-4X(G_{4X}+XG_{4XX})-4X^2(5A_3 +
2XA_{3X})\right]^{-1},
\end{align}
and the dimensionless parameters
\begin{align}
\Upsilon_1&:=-\frac{4X^2A_3^2}{G_4(G_{4XX}+2A_3+XA_{3X})+G_{4X}(G_{4X}+XA_3)},
\\
\Upsilon_2&:=-\frac{4XA_3(G_{4X}+2XG_{4XX}+5XA_3+2X^2A_{3X})}%
{5[G_4(G_{4XX}+2A_3+XA_{3X})+G_{4X}(G_{4X}+XA_3)]}.
\end{align}
These two parameters characterize the deviation from the Horndeski theory.
In terms of more familiar quantities,
Eqs.~\eqref{solydhost} and~\eqref{solzdhost} are written as
\begin{align}
\Phi'=G_N \left(\frac{{\cal M}}{r^2}+\frac{\Upsilon_1{\cal M}''}{4}\right),
\label{VB:Phi}
\\
\Psi'=G_N \left(\frac{{\cal M}}{r^2}-\frac{5\Upsilon_2{\cal M}'}{4r}\right).
\label{VB:Psi}
\end{align}
From Eqs.~\eqref{VB:Phi} and~\eqref{VB:Psi} we see the followings:
(i) the Vainshtein mechanism works outside a source because
${\cal M}=\,$const there; (ii) the Vainshtein mechanism breaks
inside a source where ${\cal M}$ is no longer constant.
This result implies that DHOST theories can be constrained
by astronomical observations of stars, galaxies, and
galaxy clusters~\cite{Koyama:2015oma,Saito:2015fza,%
Sakstein:2015zoa,Sakstein:2015aac,Jain:2015edg,%
Sakstein:2016ggl,Sakstein:2016lyj,Salzano:2017qac,Saltas:2018mxc}.
This class of gravity modification
can even be tested through
the speed of sound in the atmosphere of the Earth~\cite{Babichev:2018rfj}.

We have thus seen that gravity is modified inside a source
in a simple DHOST theory.
Such partial breaking of Vainshtein screening occurs in more
general DHOST theories as well. Of particular interest
are theories satisfying $c_{\rm GW}^2=1$ (i.e., theories
satisfying Eqs.~\eqref{c1t1}--\eqref{c1t3}).
After some tedious calculations one ends up
with~\cite{Crisostomi:2017lbg,Langlois:2017dyl,Dima:2017pwp,Crisostomi:2017pjs}
\begin{align}
  \Phi'&=G_N \left(\frac{{\cal M}}{r^2}+\frac{\Upsilon_1{\cal M}''}{4}\right),
  \label{VB:Phi2}
  \\
  \Psi'&=G_N \left(\frac{{\cal M}}{r^2}
  -\frac{5\Upsilon_2{\cal M}'}{4r}+\Upsilon_3{\cal M}''
  \right),
  \label{VB:Psi2}
\end{align}
where
\begin{align}
8\pi G_N:=\left[2(f-Xf_X-3X^2A_3)\right]^{-1},
\end{align}
and
\begin{align}
\Upsilon_1:=-\frac{(f_X-XA_3)^2}{A_3 f},
\quad
\Upsilon_2:=\frac{8Xf_X}{5f},
\quad
\Upsilon_3:=\frac{(f_X-XA_3)(f_X+XA_3)}{4A_3f}.
\end{align}
The three parameters are not independent:
$2\Upsilon_1^2-5\Upsilon_1\Upsilon_2-32\Upsilon_3^2=0$.
Note that in deriving the above result we have implicitly assumed
that $A_3\neq 0$, which means that
we need to be more careful when considering DHOST theories
devoid of the decay of gravitational waves into $\phi$~\cite{Creminelli:2018xsv}.
The Vainshtein regime of this special class of DHOST theories
have been investigated recently in~\cite{Hirano:2019scf,Crisostomi:2019yfo}.

Going beyond the weak gravity regime,
relativistic stars in DHOST theories have been
studied in~\cite{Babichev:2016jom,Sakstein:2016oel,Chagoya:2018lmv,Kobayashi:2018xvr}.

As explained in the previous section,
class Ia DHOST theories can be mapped to the Horndeski theory
via a disformal transformation. Therefore, DHOST theories
with minimally coupled matter
are equivalent to the Horndeski theory with
disformally coupled matter.
This is the reason why the behavior of gravity in DHOST theories
is different from that in the Horndeski theory
in the presence of matter.

\section{Black holes in Horndeski theory and beyond}\label{sec:BHs}

In general relativity, a black hole is characterized solely by
its mass, angular momentum, and electric charge.
This is the well-known no-hair theorem.
In scalar-tensor theories,
the scalar field would not be regular at the horizon
in many cases unless it has a trivial profile.
The no-hair theorem can thus be extended to cover a wider class of
theories~\cite{Herdeiro:2015waa,Volkov:2016ehx}, though
it can certainly be evaded, e.g., by a nonminimal coupling
to the Gauss-Bonnet term~\cite{Kanti:1995vq,Antoniou:2017acq,Antoniou:2017hxj,Bakopoulos:2018nui}.
In light of the modern reformulation of the Horndeski theory,
it has been argued that nontrivial profiles of the Galileon field
are not allowed around static and spherically symmetric black holes~\cite{Hui:2012qt}.
The proof of~\cite{Hui:2012qt} is based on the shift symmetry of the scalar field and
several other assumptions.
It is therefore intriguing to explore how one can circumvent
the no-hair theorem in the context of the Horndeski/beyond Horndeski theories.

For example, by tuning the form of the Horndeski functions
one can evade the no-hair theorem~\cite{Sotiriou:2013qea,Sotiriou:2014pfa}.
Another possibility is relaxing the assumptions on
the time independence of $\phi$ and/or its asymptotic
behavior~\cite{Rinaldi:2012vy,Anabalon:2013oea,%
Minamitsuji:2013ura,Babichev:2013cya,Cisterna:2014nua}.
In particular, it is important to notice that
in shift-symmetric scalar-tensor theories
the metric can be static even if the scalar field is linearly dependent on
time~\cite{Babichev:2010kj},
\begin{align}
\D s^2 &= -h(r)\D s^2+\frac{\D r^2}{f(r)}+r^2
\left(\D\theta^2+\sin^2\theta\D\varphi^2\right),
\label{BH:ansatz_met}
\\
\phi&=qt +\psi(r),\quad q={\rm const},\label{BH:ansatz_sc}
\end{align}
because the field equations depend on $\phi$
through $\partial_\mu\phi$ due to the shift symmetry.
Starting from the ansatz~\eqref{BH:ansatz_met} and~\eqref{BH:ansatz_sc},
various hairy black hole solutions have been obtained
in scalar-tensor theories with the derivative coupling of the form
$\sim G^{\mu\nu}\phi_\mu\phi_\nu$ in~\cite{Babichev:2013cya}.
The same strategy was then used to derive hairy black hole solutions
from more general Lagrangians in the
Horndeski family~\cite{Kobayashi:2014eva,Babichev:2016fbg,Tretyakova:2017lyg},
its bi-scalar extension~\cite{Charmousis:2014zaa},
and the GLPV/DHOST theories~\cite{Babichev:2016kdt,Babichev:2017guv,BenAchour:2018dap,%
Motohashi:2019jre} (see~\cite{Babichev:2016rlq,Lehebel:2018zga} for a review).
Some of these solutions have the Schwarzschild(-(A)dS) geometry
dressed with nontrivial scalar-field profiles, i.e., a stealth property.

As explained in the previous section,
strong constraints have been imposed on
scalar-tensor theories
as alternatives to dark energy after GW170817.
Implications of the limit $c_{\rm GW}^2=1$ for black holes
in scalar-tensor theories have been discussed
in~\cite{Babichev:2017lmw,Tattersall:2018map,BenAchour:2018dap}.

Perturbations of black holes in scalar-tensor theories
are also worth investigating for the same reasons as
in the case of cosmological perturbations:
one can judge the stability of a given black hole solution
and give predictions for observations.
As in general relativity,
for a spherically symmetric background it is convenient to
decompose metric perturbations into even parity (polar)
and odd parity (axial) modes.
The scalar field perturbations come into play only in
the even parity sector.
Since the Horndeski theory and its extensions
preserve parity, the equations of motion for the even and odd modes
are decoupled.
Within the Horndeski theory,
the quadratic actions and the stability conditions
of the even and odd parity perturbations
were derived for a general static and spherically symmetric
background with a time-independent scalar field
in~\cite{Kobayashi:2012kh,Kobayashi:2014wsa}.\footnote{These papers contain
typos, some of which were pointed out in~\cite{Ganguly:2017ort,Mironov:2018uou}.
The reader is recommended to refer to the latest versions
of {\ttfamily arXiv:1202.4893} and {\ttfamily arXiv:1402.6740}.}
See also~\cite{Cisterna:2015uya,Tattersall:2018nve}.
General black hole perturbation theories covering a wider class of
Lagrangians have been developed in~\cite{Kase:2014baa,Tattersall:2017erk}.
Odd parity perturbations and
the stability of static and spherically symmetric solutions
with a linearly time-dependent scalar field are discussed
in~\cite{Ogawa:2015pea,Takahashi:2016dnv}, but their
conclusions about instabilities have been
questioned~\cite{Babichev:2018uiw}.

The perturbation analysis of spherically symmetric solutions
in the Horndeski theory can be applied not only to black holes,
but also to wormholes.
The structure of the stability conditions
for spherically symmetric solutions
is analogous to that for cosmological solutions,
which allows us to formulate the no-go theorem for stable wormholes
in the Horndeski theory in a similar way to proving the no-go
for nonsingular cosmologies introduced in
Sec.~\ref{sec:nogotheo}~\cite{Rubakov:2015gza,Rubakov:2016zah,Kolevatov:2016ppi,Evseev:2017jek}.
Also in the wormhole case, theories beyond Horndeski admit
stable solutions~\cite{Franciolini:2018aad,Mironov:2018uou}.

\section{Conclusion}\label{sec:final}

In this review, we have discussed recent advances in
the Horndeski theory~\cite{Horndeski:1974wa} and its healthy extensions,
i.e., degenerate higher-order scalar-tensor (DHOST) theories~\cite{Langlois:2015cwa}.
We have reviewed how the Horndeski theory was ``rediscovered''
in its modern form in the course of developing the
Galileon theories~\cite{Deffayet:2011gz,Kobayashi:2011nu}.
This rediscovery has stimulated extensive researches
on physics beyond the cosmological standard model
using the general framework of scalar-tensor theories.
Along with renewed interest in the Horndeski theory,
the border of Ostrogradsky-stable scalar-tensor theories
has expanded recently to include more general DHOST theories.
Among DHOST theories, it is quite likely that only
the Horndeski theory and its disformal relatives admit
stable cosmological solutions and hence can potentially
be viable~\cite{Langlois:2017mxy,deRham:2016wji}.

In light of GW170817, we have seen that some of free functions
in DHOST Lagrangians can be strongly constrained
upon imposing $c_{\rm GW}=1$~\cite{Creminelli:2017sry,Sakstein:2017xjx,%
Ezquiaga:2017ekz,Baker:2017hug,%
Bartolo:2017ibw,Kase:2018iwp,Ezquiaga:2018btd,Kase:2018aps}.
However, one must be careful
about the range of the validity of modified gravity under consideration.
The cutoff scale of modified gravity
as an alternative to dark energy
may be close to the energy scales
observed at LIGO~\cite{deRham:2018red}, and
modified gravity in the early universe (i.e., at much higher energies)
is free from the constraint $c_{\rm GW}\simeq 1$.
Even if one imposes $c_{\rm GW}=1$, there still is
an interesting class of scalar-tensor theories,
in which the nonstandard behavior of gravity arises only
inside matter~\cite{Crisostomi:2017lbg,Langlois:2017dyl,Dima:2017pwp,Crisostomi:2017pjs}.

Having obtained a general framework of healthy scalar-tensor theories,
it would be exciting to test gravity with
cosmological and astrophysical observations
as well as to explore novel models of the early universe.
Now we are at the dawn of gravitational-wave astrophysics and cosmology, and
gravitational waves allow us to access physics
at extremely high energies and in the strong-gravity regime.
In view of this,
we hope that the general framework presented in this review will
prove more and more useful in exploring the fundamental nature of gravity.
We also hope that generalizing gravity will result in
gaining yet deeper insights into theoretical aspects of
gravity.

\section*{Acknowledgements}
I am grateful to
Shingo Akama, Yuji Akita, Antonio De Felice,
Takashi Hiramatsu, Shin'ichi Hirano, Aya Iyonaga,
Xian Gao, Kohei Kamada, Rampei Kimura, Taro Kunimitsu,
Hayato Motohashi,
Tatsuya Narikawa, Sakine Nishi, Atsushi Nishizawa,
Hiromu Ogawa, Seiju Ohashi, Ryo Saito, Maresuke Shiraishi, Teruaki Suyama,
Hiroaki W. H. Tahara, Kazufumi Takahashi, Tomo Takahashi, Yu-ichi Takamizu,
Norihiro Tanahashi, Hiroyuki Tashiro, Shinji Tsujikawa,
Yuki Watanabe,
Kazuhiro Yamamoto, Masahide Yamaguchi,
Daisuke Yamauchi, Jun'ichi Yokoyama, and Shuichiro Yokoyama
for fruitful collaborations on the Horndeski theory and modified gravity
over the recent years.
The work of TK was supported by
MEXT KAKENHI Grant Nos.~JP15H05888, JP16K17707, JP17H06359, JP18H04355,
and MEXT-Supported Program for the Strategic Research Foundation at Private Universities,
2014-2018 (S1411024).


\section*{References}

\bibliography{review_Horndeski}
\bibliographystyle{JHEP}
\end{document}